\documentclass[titlepage,pallis,twoside]{wpallis}
\usepackage{fancyh}
\usepackage[l]{floatflt}
\usepackage{epsfig}
\usepackage{amssymb}
\usepackage{latexsym}
\usepackage{times}
\usepackage{slashed,cite}
\usepackage{upgreek}
\numberwithin{equation}{section}

\usepackage{amsmath}
\usepackage{amsfonts}
\usepackage{amsbsy}
\usepackage{amscd}
\usepackage{bbm}

\newcommand{\bal}{\begin{align}}
\newcommand{\eal}{\end{align}}
\newcommand{\beqs}{\begin{subequations}}
\newcommand{\eeqs}{\end{subequations}}
\newcommand{\ec}{\end{center}}
\newcommand{\bec}{\begin{center}}

\newcommand{\eem}{\end{matrix}}
\newcommand{\bem}{\begin{matrix}}
\newcommand{\eeq}{\end{equation}}
\newcommand{\beq}{\begin{equation}}
\newcommand{\ba}{\begin{array}}
\newcommand{\ea}{\end{array}}
\newcommand{\bea}{\begin{eqnarray}}
\newcommand{\eea}{\end{eqnarray}}
\newcommand{\baq}{\begin{eqnarray}}
\newcommand{\eaq}{\end{eqnarray}}

\newcommand\eqs[2]{Eqs.~(\ref{#1}) and (\ref{#2})}

\newcommand\eqss[3]{Eqs.~(\ref{#1}), (\ref{#2}) and (\ref{#3})}

\newcommand{\ftn}{\footnotesize}

\newcommand{\ssz}{\scriptsize}
\newcommand{\TeV}{{\mbox{\rm TeV}}}

\newcommand{\GeV}{{\mbox{\rm GeV}}}

\newcommand{\sFref}[2]{Fig.~\ref{#1}-{\ftn\sf ({#2})}}
\newcommand{\sEref}[2]{Eq.~(\ref{#1}{\ftn\sf {#2}})}

\newcommand{\etal}{{\it et al.\/}}

\def\to{\rightarrow}
\def\dg{\dagger}
\def\llgm{\left\lgroup}
\def\rrgm{\right\rgroup}
\def\lf{\left(}
\def\rg{\right)}
\newcommand\vev[1]{\langle {#1} \rangle}
\newcommand{\Gr}{\ensuremath{\widetilde{G}}}
\newcommand{\Yb}{\ensuremath{Y_{B}}}
\newcommand{\Yg}{\ensuremath{Y_{\Gr}}}

\newcommand{\Vjhi}{\ensuremath{V_{\rm HI}}}
\newcommand{\Vhi}{\ensuremath{\widehat V_{\rm HI}}}
\newcommand{\Hhi}{\ensuremath{\widehat H_{\rm HI}}}
\newcommand{\Ohi}{\ensuremath{\Omega}}
\newcommand{\Omg}{\ensuremath{\Omega}}
\newcommand{\Khi}{\ensuremath{K}}
\newcommand{\Shi}{\ensuremath{S_{\rm HI}}}
\newcommand{\Vhio}{\ensuremath{\widehat V_{\rm HI0}}}

\newcommand{\mP}{\ensuremath{m_{\rm P}}}
\newcommand{\Mpq}{\ensuremath{M_{\rm PS}}}
\newcommand{\mpq}{\ensuremath{m_{\rm PS}}}
\newcommand{\Mgut}{\ensuremath{M_{\rm GUT}}}
\newcommand{\Ggut}{\ensuremath{G_{\rm PS}}}
\newcommand{\Ms}{\ensuremath{M_{\rm S}}}

\newcommand\lvc[3]{\ensuremath{\varepsilon_{{#1}{#2}{#3}}}}
\def\openep{\leavevmode\hbox{\normalsize{\boldmath $\varepsilon$}}}
\def\openone{\leavevmode\hbox{\small1\kern-3.8pt\normalsize1}}

\newcommand{\ca}{\ensuremath{{\rm a}}}
\newcommand{\ck}{\ensuremath{c_\mathcal{R}}}
\newcommand{\kx}{\ensuremath{k_S}}
\newcommand{\kn}{\ensuremath{k_{H}}}

\newcommand{\Gsm}{\ensuremath{G_{\rm SM}}}

\newcommand{\Gsn}{\ensuremath{\Gamma_{\rm I}}}
\newcommand{\GNsn}{\ensuremath{\Gamma_{\rm I\nu^c}}}
\newcommand{\Ghsn}{\ensuremath{\Gamma_{{\rm I}y}}}

\newcommand{\msn}{\ensuremath{m_{\rm I}}}

\newcommand{\deff}{\ensuremath{\delta_{\rm eff}}}

\newcommand{\hd}{{\ensuremath{H_d}}}
\newcommand{\hu}{{\ensuremath{H_u}}}

\newcommand{\ns}{\ensuremath{n_{\rm s}}}
\newcommand{\as}{\ensuremath{\alpha_{\rm s}}}

\newcommand{\rcc}{\ensuremath{\mathcal{R}}}
\newcommand{\rce}{\ensuremath{\widehat{\mathcal{R}}}}
\newcommand{\Ve}{\ensuremath{\widehat{V}}}
\newcommand{\He}{\ensuremath{\widehat{H}}}
\newcommand{\Ne}{\ensuremath{\widehat{N}}}
\newcommand{\sni}{\ensuremath{\nu^c_i}}

\newcommand{\mrh[1]}{\ensuremath{M_{#1\what\nu^c}}}
\newcommand{\mD[1]}{\ensuremath{m_{#1\rm D}}}

\newcommand{\mntau}{\ensuremath{m_{\nu_\tau}}}
\newcommand{\mnmu}{\ensuremath{m_{\nu_\mu}}}
\newcommand{\cth}{\ensuremath{c_\vartheta}}
\newcommand{\sth}{\ensuremath{s_\vartheta}}

\newcommand{\snH}{\ensuremath{\nu^c_H}}
\newcommand{\snHb}{\ensuremath{\bar\nu^c_H}}
\newcommand{\uH}{\ensuremath{u^c_H}}
\newcommand{\uHb}{\ensuremath{\bar u^c_H}}
\newcommand{\dH}{\ensuremath{d^c_H}}
\newcommand{\dHb}{\ensuremath{\bar d^c_H}}
\newcommand{\eH}{\ensuremath{e^c_H}}
\newcommand{\eHb}{\ensuremath{\bar e^c_H}}
\newcommand{\lH}{\ensuremath{\lambda_H}}
\newcommand{\lHb}{\ensuremath{\lambda_{\bar H}}}
\newcommand{\Hcc}{\ensuremath{H^c}}
\newcommand{\bHc}{\ensuremath{\bar H^c}}
\newcommand{\what}{\ensuremath{\widehat}}
\newcommand{\ovl}{\ensuremath{\overline}}

\def\ve{\varepsilon}

\def\bbet{{\bar\beta}}
\def\al{{\alpha}}
\def\bt{{\beta}}

\def\th{{\theta}}
\def\thb{{\bar\theta}}
\def\thn{{\theta_{\nu}}}

\newcommand{\Trh}{\ensuremath{T_{\rm rh}}}
\newcommand{\sg}{\ensuremath{h}}

\newcommand{\sgf}{\ensuremath{h_{\rm f}}}
\newcommand{\xsg}{\ensuremath{x_{h}}}

\newcommand{\dth}{\ensuremath{\delta h}}
\newcommand{\ld}{\ensuremath{\lambda}}
\newcommand{\ldu}{\ensuremath{\uplambda}}
\newcommand{\Ld}{\ensuremath{\Lambda}}

\newcommand{\se}{\ensuremath{\widehat h}}
\newcommand{\sex}{\ensuremath{\widehat{h}_*}}

\newcommand{\geu}{\ensuremath{\widehat g}}

\newcommand{\mef}{\ensuremath{m_{\rm eff}}}

\newcommand{\eL}{{\ensuremath{\varepsilon_{L}}}}

\def\Ka{K\"{a}hler potential}

\def\FHI{non-MHI~}

\newcommand{\tr}{{\mbox{\sf\ssz T}}}
\newcommand{\trc}{{\mbox{\sf\ssz T}}}
\newcommand{\Tr}{\mbox{\sf Tr}}
\newcommand{\diag}{\mbox{\sf\ftn diag}}

\newcommand{\Fci}{\mbox{
$\llgm\bem q^c_{i1} \cr q^c_{i2} \cr q^c_{i3}\cr l^c_i\eem\rrgm$}}

\newcommand{\Fi}{\mbox{
$\llgm\bem q_{i1}&q_{i2}&q_{i3}&l_i\eem\rrgm$}}

\newcommand{\Hcb}{\mbox{
$\llgm\bem q^c_{H1} \cr  q^c_{H2} \cr q^c_{H3}\cr
l^c_H\eem\rrgm$}}

\newcommand{\Hca}{\mbox{
$\llgm\bem \bar q^c_{H1}&\bar q^c_{H2}&\bar q^c_{H3}&\bar
l^c_H\eem\rrgm$}}

\newcommand{\bdh}{\mbox{
$\llgm\bem H^+_u & H^0_d \cr H^0_u & H^-_d\eem\rrgm$}}

\newcommand{\gmt}{\mbox{
$\llgm\bem \varepsilon_{\rm abc}g^c_{\rm c}  &\bar  g^c_\ca \cr
-\bar g^c_\ca & 0\eem\rrgm$}}

\newcommand{\dgmt}{\mbox{
$\llgm\bem \varepsilon_{\rm abc}\bar g^c_{\rm c}  &g^c_\ca \cr
-g^c_\ca & 0\eem\rrgm$}}

\newcommand\mtt[4]{\mbox{
$\llgm\bem #1 &#2 \cr #3& #4\eem\rrgm$}}

\newcommand{\bdhh}{{\ensuremath{\normalsize I{\kern-2.9pt H}}}}

\newcommand\lin[2]{\mbox{$\llgm\bem #1& #2\eem\rrgm$}}

\newcommand\stl[2]{\mbox{$\llgm\bem #1\cr #2\eem\rrgm$}}

\begin{document}

\thispagestyle{empty}

\title[]{\Large\bfseries\scshape Non-Minimal Higgs Inflation and
non-Thermal Leptogenesis in a Supersymmetric Pati-Salam Model}

\author{\large\bfseries\scshape C. Pallis and N. Toumbas}
\address[] {\sl Department of Physics, University of Cyprus, \\
P.O. Box 20537, CY-1678 Nicosia, CYPRUS}

\begin{abstract}{{\bfseries\scshape Abstract} \\
\par We consider a supersymmetric (SUSY) Grand Unified Theory (GUT)
based on the gauge group $G_{\rm PS}=SU(4)_{\rm C} \times
SU(2)_{\rm L}\times SU(2)_{\rm R}$, which incorporates non-minimal
chaotic inflation, driven by a quartic potential associated with
the Higgs fields involved in the spontaneous breaking of $\Ggut$.
The inflationary model relies on renormalizable superpotential
terms and does not lead to overproduction of magnetic monopoles.
It is largely independent of the one-loop radiative corrections
and can become consistent with the current observational data on
the inflationary observables, with the symmetry breaking scale of
$G_{\rm PS}$ assuming its SUSY value. Within our model, the strong
CP and the $\mu$ problems of the minimal SUSY standard model can
be resolved via a Peccei-Quinn symmetry. Moreover baryogenesis
occurs via non-thermal leptogenesis realized by the
out-of-equilibrium decay of the right-handed neutrinos, which are
produced by the inflaton's decay. We consider two versions of such
a scenario, assuming that the inflaton can decay to the lightest
or to the next-to-lightest right-handed neutrino. Both scenaria
can become compatible with the constraints arising from the baryon
asymmetry of the universe, the gravitino limit on the reheating
temperature and the upper bound on the light neutrino masses,
provided that the gravitino is somehow heavy. In the second
scenario, extra restrictions from the $SU(4)_{\rm C}$ GUT symmetry
on the heaviest Dirac neutrino mass and the data on the
atmospheric neutrino oscillations can be also met. }
\end{abstract} \maketitle


\setcounter{page}{1} \pagestyle{fancyplain}


\rhead[\fancyplain{}{ \bf \thepage}]{\fancyplain{}{\sl Non-MHI \&
non-Thermal Leptogenesis in a SUSY PS Model}}
\lhead[\fancyplain{}{\sl \leftmark}]{\fancyplain{}{\bf \thepage}}
\cfoot{}

\tableofcontents\vskip-1.3cm\noindent\rule\textwidth{.4pt}\\

\section{Introduction}\label{intro}

\emph{Non-minimal Higgs inflation} (non-MHI)
\cite{sm1,unitarizing, linde1} -- see also \cref{SusyHiggs} -- is
an inflationary model of chaotic type which arises in the presence
of a non-minimal coupling between a Higgs-inflaton field and the
Ricci scalar curvature, $\rcc$. It has been shown that non-MHI
based on a quartic potential with a quadratic non-minimal coupling
to gravity can be realized in both a non-supersymmetric \cite{sm1}
and a \emph{sypersymmetric} (SUSY) framework \cite{linde1},
provided that the coupling of the inflaton to $\rcc$ is strong
enough. In most of the existing models, the inflaton is identified
with the Higgs field(s) of the \emph{Standard Model} (SM) or the
next-to-MSSM (\emph{Minimal SUSY SM}) \cite{linde1} -- for
non-minimal inflation driven by an inflaton other than the Higgs
field see \cref{nmchaotic, wmap3, love, nmi, ld} for non-SUSY
models and \cref{linde2, nmN, nmN1} for SUSY ones.

On the other hand, SUSY GUTs arise as natural extensions of
Physics beyond the MSSM. Within their framework, a number of
challenges -- such as the $\mu$ problem, the generation of the
observed \emph{baryon asymmetry of the universe} (BAU) and the
existence of tiny but non-zero neutrino masses -- which the MSSM
fails to address can be beautifully arranged. The achievement of
gauge coupling unification within the MSSM suggests that the
breaking of the GUT gauge symmetry group down to the SM one,
${G_{\rm SM}}= SU(3)_{\rm C}\times SU(2)_{\rm L}\times U(1)_{Y}$,
occurs at a scale $\Mgut\simeq2\cdot10^{16}~\GeV$ through some
Higgs superfields. Therefore, the latter arise naturally as
candidates for driving non-MHI -- for earlier attempts within
non-SUSY $SU(5)$ GUT see \cref{Cervantes}. In a such situation the
GUT gauge group is already spontaneously broken during non-MHI
through the non-zero values acquired by the relevant Higgs fields.
Consequently, non-MHI does not lead to the production of
topological defects. Moreover, the potential of non-MHI possesses
a non-zero classical inclination and so, the inflationary dynamics
is largely independent of the radiative corrections. As a
consequence, the \emph{vacuum expectation values} (v.e.vs) which
the Higgs fields acquire at the end of non-MHI can be exactly
equal to the values required by the unification of the MSSM gauge
couplings. Finally, the predicted inflationary observables are
consistent with the fitting \cite{wmap} of the seven-year data of
the \emph{Wilkinson Microwave Anisotropy Probe Satellite} (WMAP7),
\emph{baryon-acoustic-oscillations} (BAO) and \emph{Hubble
constant} ($H_0$) data.

These features are to be contrasted with the widely adopted models
of standard SUSY \emph{hybrid inflation} (HI) \cite{susyhybrid},
where the spontaneous breaking of the GUT gauge symmetry takes
place at the end of HI and, thus, topological defects are
copiously formed \cite{smooth} if they are predicted by the
symmetry breaking. This is because, the standard SUSY HI is
typically driven by a singlet field whereas the Higgs fields are
confined to zero where the GUT symmetry is unbroken. Avoidance of
cosmologically disastrous topological defects can be obtained
within smooth \cite{smooth, axilleas} or shifted \cite{jean,
newshifted} HI by using either non-renormalizable \cite{smooth,
jean} or renormalizable \cite{newshifted, axilleas} superpotential
terms, which generate stable inflationary trajectories with
non-zero values for the Higgs fields. Some of the latter
constructions, though, are much more complicated than the simplest
original model \cite{susyhybrid}. In the cases of standard
\cite{susyhybrid} and shifted \cite{jean, newshifted} HI,
radiative corrections play an important role in creating the slope
of the inflationary potential and the v.e.vs of the Higgs fields
turn out to be mostly lower than the GUT SUSY symmetry scale,
since the relevant mass scale is constrained by the normalization
of the curvature perturbation \cite{wmap}. Finally, all types of
HI suffer from the problem of an enhanced (scalar) spectral index,
$\ns$, which turns out to be, mostly, well above the current data
\cite{wmap}. For several proposals aiming to improve on the latter
shortcoming of SUSY HI see \cref{gpp, battye, mhi, hinova, news}.

In this paper we present a model of non-MHI, adopting a SUSY GUT
model based on the \emph{Pati-Salam} (PS) gauge group $G_{\rm
PS}=SU(4)_{\rm C}\times SU(2)_{\rm L}\times SU(2)_{\rm R}$. Note
that SUSY PS GUT models are motivated \cite{branes} from
recent D-brane constructions and can also arise \cite{leontaris}
from the standard weakly coupled heterotic string. Employing only
renormalizable superpotential terms, we then show that the model
naturally leads to non-MHI within SUGRA avoiding thereby the
overproduction of unwanted monopoles. Also the inflationary
observables turn out to lie within the current data \cite{wmap}.
Our model possesses a number of other interesting features too.
The $\mu$-problem of the MSSM can be solved \cite{rsym} via a Peccei-Quinn
(PQ) symmetry, which also solves the strong CP problem, and the
proton is practically stable. Light neutrinos acquire masses by
the seesaw mechanism \cite{seesaw} and the BAU can be generated
through primordial non-thermal \cite{inlept} leptogenesis. We
single out two cases according to whether the inflaton decays to
the lightest \cite{baryo} or the next-to-lightest \cite{vlachos}
\emph{right-handed} (RH) neutrino. In both cases the constraints
arising from the gravitino ($\Gr$) limit \cite{gravitino, brand,
kohri} on the reheating temperature and the BAU can be met
provided that the masses of $\Gr$ lie in the multi-$\TeV$ region.
On the other hand, the second scenario gives us the opportunity to
combine the calculation of BAU with the present neutrino data
\cite{Expneutrino} and the prediction of $\Ggut$ for the masses of
the fermions of the third generation.

The plan of this paper is as follows. We present the basic
ingredients -- particle content and structure of the
superpotential and the \Ka -- of our model in Sec.~\ref{fhim}. In
Sec.~\ref{fhi} we describe the inflationary potential, derive the
inflationary observables and confront them with observations. In
Sec.~\ref{pfhi} we outline the two scenaria of non-thermal
leptogenesis, exhibit the relevant imposed constraints and
restrict the parameters of our model for each scenario. Our
conclusions are summarized in Sec.~\ref{con}. Details concerning
the derivation of the mass spectrum of the theory during \FHI are
arranged in Appendix A whereas effects of instant preheating
potentially important for some values of the parameters are
discussed in Appendix B. Throughout the text, we use natural units
for Planck's and Boltzmann's constants and the speed of light
($\hbar=c=k_{\rm B}=1$); the subscript of type $,\chi$ denotes
derivation \emph{with respect to} (w.r.t) the field $\chi$ (e.g.,
$_{,\chi\chi}=\partial^2/\partial\chi^2$); charge conjugation is
denoted by a star and $\log~[\ln]$ stands for logarithm with basis
$10~[e]$.

\section{The Pati-Salam SUSY GUT Model}\label{fhim}

We focus on a SUSY PS GUT model described in detail in
Ref.~\cite{jean} -- see also Ref.~\cite{lpnova}. The
representations and the transformation properties of the various
superfields contained in the model under $G_{\rm PS}$, their
decomposition under ${G_{\rm SM}}$, as well as their extra global
charges are presented in Table~\ref{tab1}. Recall that, in the PS
GUT models, the SM hypercharge $Q_Y$ is identified as the linear
combination $Q_Y=Q_{T^3_{\rm R}}+Q_{(B-L)}/2$ where $Q_{T^3_{\rm
R}}$ is the $SU(2)_{\rm R}$ charge generated by $T^3_{\rm
R}=\diag\lf1,-1\rg/2$ and the $Q_{(B-L)}$ is the $SU(4)_{\rm C}$
charge generated by $T^{15}_{\rm C}=\diag\lf
1,1,1,-3\rg/2\sqrt{6}$. Here $T_{\rm R}^m$ with $m=1,2,3$ are the
3 generators of $SU(2)_{\rm R}$ and $T_{\rm C}^a$ with
$a=1,...,15$ are the 15 generators of $SU(4)_{\rm C}$ with
normalizations $\Tr\lf T_{\rm C}^aT_{\rm C}^b\rg=\delta^{ab}/2$
and $\Tr\lf T_{\rm R}^mT_{\rm R}^k\rg=\delta^{mk}/2$, where $\Tr$
denotes trace of a matrix.

The $i$th generation $(i=1,2,3)$ \emph{left-handed} (LH) quark
[lepton] superfields, $u_{i\ca}$ and $d_{i\ca}$ -- where
$\ca=1,2,3$ is a color index -- [$e_i$ and $\nu_i$] are
accommodated in a superfield $F_i$. The LH antiquark [antilepton]
superfields $u^c_{i\ca}$ and $d_{i\ca}^c$ [$e^c_i$ and $\sni$] are
arranged in another superfield $F^c_i$. These can be represented
as
\bea \nonumber
&&F_i=\Fi\>\>\>\mbox{and}\>\>\>F^c_i=\Fci\>\>\>\mbox{with}\>\>\>\\
&& q_{i\ca}=\lin{d_{i\ca}}{-u_{i\ca}},
\>l_{i}=\lin{e_{i}}{-\nu_{i}},\>
q^c_{i\ca}=\stl{-u^c_{i\ca}}{d^c_{i\ca}}
\>\>\>\mbox{and}\>\>\>l^c_{i}=\stl{-\nu^c_{i}}{e^c_{i}}. \label{Fg}
\eea
The gauge symmetry $G_{\rm PS}$ can be spontaneously broken down
to $G_{\rm SM}$ through the v.e.vs which the superfields
\bea \nonumber && H^c=\Hcb\>\>\>\mbox{and}\>\>\>\bar
H^c=\Hca\>\>\>\mbox{with}\>\>\>
\\ && q^c_{ H\ca}=\stl{ u^c_{H\ca}}{d^c_{H\ca}},\>
 l^c_{H}=\stl{\nu^c_{H}}{e^c_{H}},\> \bar q^c_{H\ca}=\lin{\bar
u^c_{H\ca}}{\bar d^c_{H\ca}} \>\>\>\mbox{and}\>\>\> \bar
l^c_{H}=\lin{\bar \nu^c_{H}}{\bar e^c_{H}}, \label{Hg} \eea
acquire in the directions $\nu^c_H$ and $\bar\nu^c_H$. The
model also contains a gauge singlet $S$, which triggers the
breaking of $G_{\rm PS}$, as well as an $SU(4)_{\rm C}$ {\bf 6}-plet $G$,
which splits into $g_\ca^c$ and $\bar{g}^c_\ca$ under $G_{SM}$
and gives \cite{leontaris} superheavy masses to $d^c_{H\ca}$ and
$\bar{d}^c_{H\ca}$. In particular, $G$ can be represented by an
antisymmetric $4\times4$ matrix as follows
\beq G=\gmt\>\Rightarrow\>\bar G=\dgmt. \eeq
Here $\bar G$ is the dual tensor of $G$, defined by $\bar
G^{IJ}=\varepsilon^{IJKL}G_{KL}$ which transforms under
$SU(4)_{\rm C}$ as $U_{\rm C}^*\bar G U_{\rm C}^\dagger$. Also
$\varepsilon^{IJKL}=\varepsilon_{IJKL}$  [$\lvc{\ca}{\rm b}{\rm
c}$] is the well-known antisymmetric tensor acting on the
$SU(4)_{\rm C}$  [$SU(3)_{\rm C}$] indices with
$\varepsilon_{1234}=1$  [$\lvc{1}{2}{3}=1$]. In the simplest
realization of this model \cite{leontaris, jean}, the electroweak
doublets $\hu$ and $\hd$, which couple to the up and down quark
superfields respectively, are exclusively contained in the
bidoublet superfield $\bdhh$, which can be written as
\beq \bdhh=\llgm\bem \hu & \hd\eem\rrgm=\bdh. \eeq

\renewcommand{\arraystretch}{1.2}

\begin{table}[!t]
\begin{center}
\begin{tabular}{|ccccccc|}
\hline {\sc Super-}&{\sc Represe-}&{\sc Trasfor-}&{\sc
Decompo-}&\multicolumn{3}{c|}{ {\sc Global} }
\\
\multicolumn{1}{|c}{\sc fields}&{\sc ntations} &{\sc mations}
&{\sc sitions} &\multicolumn{3}{c|}{ {\sc Charges}}
\\
\multicolumn{1}{|c}{}&{\sc under $G_{\rm PS}$} &{\sc under $G_{\rm
PS}$}&{\sc under $G_{\rm SM}$} &{$R$} &{PQ} &{$\mathbb{Z}^{\rm
mp}_2$}
\\\hline \hline
\multicolumn{7}{|c|}{\sc Matter Superfields}
\\ \hline
{$F_i$} &{$({\bf 4, 2, 1})$}&$F_iU_{\rm L}^{\dagger}U^\trc_{\rm
C}$ & $Q_{i\rm a}({\bf 3,2},1/6)$&$1$ &
$-1$ &$1$\\
&&& $L_i({\bf 1,2},-1/2)$&&&\\
{$F^c_i$} & {$({\bf \bar 4, 1, 2})$}&$U_{\rm C}^\ast U_{\rm
R}^\ast
F^c_i$ &$u_{i\rm a}^c({\bf\bar 3,1},-2/3)$&{ $1$ }&{$0$}&{$-1$}\\
&&& $d_{i\rm a}^c({\bf\bar 3,1},1/3)$&&&\\
&&& $\nu_i^c({\bf 1,1},0)$&&&\\
&&& $e_i^c({\bf 1,1},1)$&&&\\\hline%
\multicolumn{7}{|c|}{\sc Higgs Superfields}\\ \hline {$H^c$}
&{$({\bf \bar 4, 1, 2})$}& $U_{\rm C}^\ast U_{\rm R}^\ast H^c$
&$u_{H{\rm a}}^c({\bf \bar 3,1},-2/3)$&{$0$}&{$0$} & {$0$}\\
&&& $d_{H{\rm a}}^c({\bf \bar 3,1},1/3)$&&&\\
&&& $\nu_H^c({\bf 1,1},0)$&&&\\
&&& $e_H^c({\bf 1,1},1)$&&&\\
{$\bar H^c$}&$({\bf 4, 1, 2})$& $\bar{H}^cU^\tr_{\rm R}
U^\trc_{\rm C}$&$\bar u_{H{\rm a}}^c({\bf 3,1},2/3)$&{$0$}&{$0$}&{$0$} \\
&&& $\bar d_{H\rm a}^c({\bf 3,1},-1/3)$&&&\\
&&& $\bar \nu_H^c({\bf 1,1},0)$&&&\\
&&& $\bar e_H^c({\bf 1,1},-1)$&&&\\
{$S$} & {$({\bf 1, 1, 1})$}&$S$ &$S ({\bf 1,1},0)$&$2$ &$0$ &$0$ \\
{$G$} & {$({\bf 6, 1, 1})$}&$U_{\rm C}GU^\trc_{\rm C}$ &$\bar
g_\ca^c({\bf 3,1},-1/3)$&$2$ &$0$ &$0$\\ &&& $g_\ca^c({\bf \bar
3,1},1/3)$&&&\\\hline
{$\bdhh$} & {$({\bf 1, 2, 2})$}&$U_{\rm L}\bdhh U^\tr_{\rm R}$ &$H_u({\bf 1,2},1/2)$&$0$ &$1$ &$0$\\
&&& $H_d({\bf 1,2}, -1/2)$&&&\\ \hline
{$P$} &{$({\bf 1, 1, 1})$}& $P$ &$P ({\bf 1,1},0)$&{$1$}&{$-1$} & {$0$} \\
{$\bar P$}&$({\bf 1, 1, 1})$& $\bar P$&$\bar P ({\bf
1,1},0)$&{$0$}&{$1$}&{$0$}\\ \hline
\end{tabular}
\end{center}
\vchcaption{\sl\ftn The representations, the transformations under
$G_{\rm PS}$, the decompositions under $G_{\rm SM}$ as well as the
extra global charges of the superfields of our model. Here $U_{\rm
C}\in SU(4)_{\rm C},~U_{\rm L}\in SU(2)_{\rm L},~U_{\rm R}\in
SU(2)_{\rm R}$ and $\tr,~\dagger$ and $\ast$ stand for the
transpose, the hermitian conjugate and the complex conjugate of a
matrix respectively.}\label{tab1}
\end{table}

In addition to $G_{\rm PS}$, the model possesses two global $U(1)$
symmetries, namely a PQ and an R symmetry, as well as a discrete
$\mathbb{Z}_2^{\rm mp}$ symmetry (`matter parity') under which
$F$, $F^c$ change sign. The last symmetry forbids undesirable
mixings of $F$ and $\bdhh$ and/or $F^c$ and $H^c$. The imposed
$U(1)$ R symmetry, $U(1)_R$, guarantees the linearity of the
superpotential w.r.t the singlet $S$. Although $S$ does not play
the role of the inflaton in our case -- in contrast to the case of
HI -- we explicitly checked that the presence of a quadratic $S^2$
term would lead to the violation of the stability of the
inflationary trajectory. Finally the $U(1)$ PQ symmetry,
$U(1)_{\rm PQ}$, assists us to generate the $\mu$-term of the
MSSM. Although this goal could be easily achieved \cite{king} by
coupling $S$ to $\bdhh^2$ and using the fact that $S$, after
gravity-mediated SUSY breaking, develops a v.e.v, we here prefer
to follow \cref{rsym, jean} imposing a PQ symmetry on the
superpotential and introducing a pair of gauge singlet superfields
$P$ and $\bar{P}$. The PQ breaking occurs at an intermediate scale
through the v.e.vs of $P$, $\bar{P}$, and the $\mu$-term is
generated via a non-renormalizable coupling of $P$ and $\bdhh$. We
do not adopt here the resolution to the $\mu$-problem suggested in
\cref{king}, since it introduces a renormalizable term which
creates {\sf (i)} a decay channel of the inflaton which leads to a
high reheating temperature in conflict with the $\Gr$ constraint
and {\sf (ii)} a tachyonic instability in the $\hu-\hd$ system
during \FHI -- as occurring in \cref{nmN}.  Lifting both
shortcomings requires an unnaturally small value for the relevant
coupling constant in our scenario (of order $10^{-6}$ or so),
which is certainly undesirable. Following \cref{jean}, we
introduce into the scheme quartic (non-renormalizable)
superpotential couplings of $\bar{H}^c$ to $F^c_i$, which generate
intermediate-scale masses for the $\sni$ and, thus, masses for the
light neutrinos, $\nu_i$, via the seesaw mechanism. Moreover,
these couplings allow for the decay of the inflaton into RH
neutrinos, $\nu^c_i$, leading to a reheating temperature
consistent with the $\Gr$ constraint with more or less natural
values of the parameters. As shown finally in \cref{jean}, the
proton turns out to be practically stable in this model.

The superpotential $W$ of our model splits into three
parts:
\beq W=W_{\rm MSSM}+W_{\rm PQ}+W_{\rm HPS}, \label{Wtotal}\eeq
where $W_{\rm MSSM}$ is the part of $W$ which contains the usual
terms -- except for the $\mu$ term -- of the MSSM, supplemented by
Yukawa interactions among the left-handed leptons and $\sni$:
\bea \nonumber  W_{\rm MSSM} = y_{ij} F_i \bdhh F^c_j= \hspace*{5.cm} \\
 = y_{ij}\lf \hd^\tr\openep {L}_ie^c_j -\hu^\tr \openep L_i\nu^c_j
+\hd^\tr \openep Q_{i\ca}d^c_{j\ca} -\hu^\tr \openep {Q}_{i\ca}
u^c_{j\ca} \rg,\>\>\>\mbox{with}\>\>\>\openep=\llgm\bem0&1\cr-1
&0\eem\rrgm\cdot \label{wmssm}\eea
Here $Q_{i\ca}=\lin{u_{i\ca}}{d_{i\ca}}^\tr$ and
$L_{i}=\lin{\nu_i}{e_i}^\tr$ are the $i$-th generation $SU(2)_{\rm
L}$ doublet LH quark and lepton superfields respectively.
Summation over repeated color and generation indices is also
assumed. Obviously the model predicts Yukawa unification at
$\Mgut$ since the third family fermion masses originate from a
unique term in the underlying GUT. It is shown \cite{oliveira,
shafi} that exact Yukawa unification combined with
non-universalities in the gaugino sector and/or the scalar sector
can become consistent with a number of phenomenological and
cosmological low-energy requirements. The present model can be
augmented \cite{lpnova} with other Higgs fields so that $\hu$ and
$\hd$ are not exclusively contained in $\bdhh$ but receive
subdominant contributions from other representations too. As a
consequence, a moderate violation of the exact Yukawa unification
can be achieved, allowing for an acceptable low-energy phenomenology,
even with universal boundary conditions for the soft SUSY breaking
terms. However, here we prefer to work with the simplest version
of the PS model.

The second term in the \emph{right hand side} (RHS) of
\Eref{Wtotal}, $W_{\rm PQ}$, is the part of $W$ which is relevant
for the spontaneous breaking of $U(1)_{\rm PQ}$ and the generation
of the $\mu$ term of the MSSM. It is given by
\beq\label{Wpq} W_{\rm PQ}=\ld_{\rm PQ} \frac{P^2
\bar{P}^2}{\Ms}-\ld_\mu \frac{P^2}{2\Ms}\Tr\left(\bdhh\openep
\bdhh^{\tr}\openep\right),  \eeq
where $\Ms\simeq 5\cdot 10^{17}~{\rm GeV}$ is the String scale.
The scalar potential, which is generated by the first term in the
RHS of \Eref{Wpq}, after gravity-mediated SUSY breaking is studied
in \cref{rsym,jean}. For a suitable choice of parameters, the
minimum lies at $\vert \vev{P}\vert = \vert \vev{\bar{P}}\vert
\sim \sqrt{m_{3/2}M_{\rm S}}$. Hence, the PQ symmetry breaking
scale is of order $\sqrt{m_{3/2} \Ms} \simeq \lf10^{10} -
10^{11}\rg~\GeV$ and the $\mu$-term of the MSSM is generated from
the second term of the RHS of \Eref{Wpq} as follows:
\beq -\ld_\mu \frac{\vev{P}^2}{2\Ms}\Tr\left(\bdhh\openep
\bdhh^{\tr}\openep\right)=\mu\hd^\tr\openep\hu\>\>\Rightarrow
\>\>\mu\simeq\ld_\mu {\vev{P}^2\over\Ms},\eeq
which is of the right magnitude if $\lambda_\mu\sim(0.001-0.01)$.
Let us note that $V_{\rm PQ}$ has an additional local minimum at
$P=\bar{P}=0$, which is separated from the global PQ minimum by a
sizable potential barrier, thus preventing transitions from
the trivial to the PQ vacuum. Since this situation persists at all
cosmic temperatures after reheating, we are obliged to assume
that, after the termination of non-MHI, the system emerges with
the appropriate combination of initial conditions so that it lies
\cite{curvaton} in the PQ vacuum.

Finally, the third term in the RHS of \Eref{Wtotal}, $W_{\rm HPS}$,
is the part of $W$ which is relevant for non-MHI, the spontaneous
breaking of $G_{\rm PS}$ and the generation of intermediate
Majorana [superheavy] masses for $\sni$ [$\dH$ and $\dHb$]. It
takes the form
\beq W_{\rm HPS}=\ld S \lf\bHc\Hcc -\Mpq^2\rg +\lH H^{c\trc} G
\openep H^c + \lHb \bar{H}^{c} \bar G \openep \bar{H}^{c\trc}+
\ld_{i\nu^c} \frac{\lf\bar{H}^c F_i^c\rg^2}{\Ms},\label{Whi} \eeq
where $\Mpq$ is a superheavy mass scale related to $\Mgut$ -- see
\Sref{fhi2}. The parameters $\ld$ and $\Mpq$ can be made positive
by field redefinitions. It is worth emphasizing that our
inflationary model is totally tied on renormalizable
superpotential terms, contrary to the model of shifted HI
\cite{jean}, where a non-renormalizable term added in the RHS of
\Eref{Whi} plays a crucial role in the inflationary dynamics.

Suppressing henceforth the color indices, we can express $W_{\rm
HPS}$ in terms of the components of the various superfields. We
find
\begin{eqnarray}
W_{\rm HPS}&=& \ld
S\lf\snH\snHb+\eH\eHb+\uH\uHb+\dH\dHb-\Mpq^2\rg\nonumber
\\&-&2\lH\lf \nu^c_H d^c_H -e^c_H u^c_H \rg\bar{g}^c+ 2\lH u^c_H d^c_H g^c\nonumber\\
&-& 2\lHb \lf\bar{\nu}^c_H\bar{d}^c_H- \bar{e}^c_H\bar{u}^c_H\rg
g^c + 2\lHb\bar{u}^c_H \bar{d}^c_H \bar{g}^c\nonumber
\\&+&\ld_{i\nu^c} \lf\eHb e_i^c+\dHb
d_i^c-\snHb\nu_i^c-\uHb u_i^c\rg^2/\Ms.\label{Wcom}
\end{eqnarray}
Let us note in passing that the combination of two [three]
color-charged objects in a term involves a contraction
of the color indices with the symmetric [antisymmetric] invariant tensor
$\delta_{\ca\rm b}~[\lvc{\ca}{\rm b}{\rm c}]$, e.g.,
$\uH\uHb=\delta_{\ca\rm b} u_{H\ca}^c \bar u_{H\rm b}^c~[u^c_H
d^c_H g^c=\lvc{\ca}{\rm b}{\rm c} u^c_{H\ca} d^c_{H\rm b}g_{\rm
c}^c]$.

According to the general recipe \cite{linde1,linde2}, the
implementation of \FHI within SUGRA requires the adoption of a
frame function, $\Ohi$, related to the K\"ahler potential, $\Khi$,
as follows
\beq  \Ohi=-3e^{-\Khi/3\mP^2}=-3+{
H^{c\dg}\Hcc\over\mP^2}+{\bHc\bar H^{c\dg}\over\mP^2}+{\Tr \lf
G^\dg G\rg \over2\mP^2}+
{|S|^2\over\mP^2}-{\kx}{|S|^4\over\mP^4}-{3\kn\over2\mP^2}\lf
\bHc\Hcc +{\rm h.c.}\rg,\label{minK}\eeq
where the complex scalar components of the superfields $\Hcc,
\bHc, G$ and $S$ are denoted by the same symbol and the
coefficients $\kx$ and $\kn$ are taken real. It is clear from
\Eref{minK} that we adopt the standard quadratic non-minimal
coupling for Higgs-inflaton, which respects the gauge and the
global symmetries of the model. We also added the fifth term in
the RHS of \Eref{minK} in order to cure the tachyonic mass problem
encountered in similar models \cite{linde1, linde2} -- see
\Sref{fhi1}. In terms of the components of the various fields,
$\Khi$ in \Eref{minK} reads
\beq \Khi=-3\mP^2\ln \lf1-
{\phi^\al\phi^{*\al}\over3\mP^2}+{\kx}{|S|^4\over3\mP^4}+
{\kn\over2\mP^2}\lf\snH\snHb+\eH\eHb+\uH\uHb+\dH\dHb+{\rm
h.c.}\rg\right) \label{Kcom} \eeq
with $\phi^\al=\snH, \snHb, \eH, \eHb, \uH, \uHb, \dH, \dHb, g^c,
\bar g^c, S$ and summation over the repeated Greek indices -- e.g.
$\al$ and $\bbet$ -- is implied

In the limit where $\mP$ tends to infinity, we can obtain the SUSY
limit, $V_{\rm SUSY}$, of the SUGRA potential, $\widehat V_{\rm
HF}$ -- see \Sref{fhi}. Assuming that the SM non-singlet
components vanish, $V_{\rm SUSY}$ turns out to be
\beq \label{VF}V_{\rm SUSY} = \ld^2 \vert\lf\snHb \snH  -
\Mpq^2\rg\vert^2 + \ld^2 \vert S \vert^2 \left(|\snH|^2
+|\snHb|^2\right). \eeq
On the other hand, assuming minimal gauge kinetic functions, the
D-term scalar potential $V_{\rm HD}$ of the PS Higgs fields takes
the form
\begin{equation}
V_{\rm HD}=\frac{g^2}{2}  \sum_{a=1}^{15}\lf \bar{H}^{c*} T_{\rm
C}^a \bar{H}^{c\trc} -{H^c}^\dagger T_{\rm C}^{a*} H^c
\rg^2+\frac{g^2}{2} \sum_{m=1}^{3}\lf \bar{H}^{c*} T_{\rm R}^m
\bar{H}^{c\tr}  -{H^c}^\dagger T_{\rm R}^{m*} H^c \rg^2,
\label{Vd}
\end{equation}
where $g$ is the (unified) gauge coupling constant of $\Ggut$. In
terms of the components of $\bHc$ and $\Hcc$, $V_{\rm HD}$ reads:
\bea \nonumber V_{\rm HD}&=&{g^2\over 8}\Big[\lf\bar\nu^{c*}_H
\eHb + \snHb\bar e_H^{c*} - \nu^{c*}_H\eH - \snH e_H^{c*}\rg^2-
\lf \snH e_H^{c*}-\nu^{c*}_H \eH+\snHb\bar
e_H^{c*}-\bar\nu^{c*}_H\eHb\rg^2\\ &+& \nonumber \lf\bar\nu^{c*}_H
\uHb + \snHb\bar u_H^{c*}-\nu^{c*}_H\uH-\snH u_H^{c*}\rg^2-\lf
\nu^{c*}_H \uH - \snH u_H^{c*}
-\bar\nu^{c*}_H\uHb +\snHb\bar u_H^{c*}\rg^2 \\
&+& {3\over2}\lf|\snH|^2 - |\snHb|^2\rg^2 + \lf|\snHb|^2 -
|\snH|^2\rg^2+\cdots\Big].\label{Vd1}\eea
Here the first line includes contributions arising from the sum
over the $SU(2)_{\rm R}$ generators $T_{\rm R}^1$ and $T_{\rm
R}^{2}$ in \Eref{Vd}, which are the well-known Pauli matrices. The
second line includes contributions from the sum over the
$SU(4)_{\rm C}$ generators $T^{7+2\ca}$ [$T^{8+2\ca}$]
$(\ca=1,2,3)$ with $1/2~[1/2]$ and $-i/2~[i/2]$ in the $\ca4$
$[4\ca]$ entries respectively and zero everywhere else, and the
third for $T_{\rm C}^{15}$ and $T_{\rm R}^3$. The ellipsis
represents terms including exclusively the SM non-singlet
directions of $\bHc$ and $\Hcc$. Vanishing of the D-terms is
achieved for $|\snHb|=|\snH|$ with the other components of $\bHc$
and $\Hcc$ frozen at zero. Restricting ourselves to the D-flat
direction, from $V_{\rm SUSY}$ in Eq.~(\ref{VF}) we find that the
SUSY vacuum lies at
\beq
\vev{S}\simeq0\>\>\>\mbox{and}\>\>\>\left|\vev{\snH}\right|=\left|\vev{\snHb}\right|=\Mpq.
\label{vevs} \eeq
Therefore, $W_{\rm HPS}$ leads to a spontaneous breaking of
$\Ggut$. The same superpotential, $W_{\rm HPS}$, also gives rise
to a stage of \FHI as analyzed in \Sref{fhi}. Indeed, along the
D-flat direction $|\snH|=|\snHb|\gg \Mpq$ and $S=0$, $V_{\rm
SUSY}$ tends to a quartic potential and so, $W_{\rm HPS}$ can be
employed in conjunction with $\Khi$ in \Eref{minK} for the
realization of \FHI along the lines of \cref{linde2}.

It should be mentioned that soft SUSY breaking and instanton
effects explicitly break $U(1)_R\times U(1)_{\rm PQ}$ to
$\mathbb{Z}_2\times \mathbb{Z}_6$. The latter symmetry is
spontaneously broken by $\vev{P}$ and $\vev{\bar{P}}$. This would
lead to a domain wall problem if the PQ transition took place after
non-MHI. However, as we already mentioned above, $U(1)_{\rm PQ}$ is
assumed already broken before or during non-MHI. The final
unbroken symmetry of the model is $\Gsm \times \mathbb{Z}_2^{\rm
mp}$.

\section{The Inflationary Scenario}\label{fhi}

We below outline the salient features of our inflationary scenario
(\Sref{fhi1}) and then, we present its predictions in
Sec.~\ref{num1}, calculating a number of observable quantities
introduced in Sec.~\ref{fhi2}.

\subsection{Structure of the Inflationary
Action}\label{fhi1}

Following the conventions of \cref{nmN}, we write the action of
our model in the \emph{Jordan frame} (JF) as follows:
\beq \label{Snpq} \Shi=\int d^4x\sqrt{-g}\lf{1\over6}\mP^2\Ohi\rcc
+\mP^2\Omega_{\al\bbet}
\partial_\mu \phi^{\al}\partial^\mu \phi^{*\bbet} -\Vjhi+\cdots\rg ,\eeq
where  $\Omg_{\al\bbet}=\Omg_{,\phi^\al\phi^\bbet}$ and $\phi^\al$
are identified below  \Eref{Kcom}. The ellipsis represents terms
arising from the covariant derivatives $D_\mu H^c$ and $D_\mu\bar
H^c$, and terms that include the on-shell auxiliary axial-vector
field \cite{linde1, nmN}, which turn out to be irrelevant for our
analysis below. Here $\Vjhi =\Ohi^2\what V_{\rm HF}/9 +V_{\rm
HD}$, with $\what V_{\rm HF}$ being the \emph{Einstein frame} (EF)
F--term SUGRA scalar potential, which is obtained from $W_{\rm
HPS}$ in Eq.~(\ref{Wcom}) -- without the last term of the RHS --
and $\Khi$ in Eq.~(\ref{Kcom}) by applying \cite{sugraL}
\begin{equation}
\what V_{\rm HF}=e^{\Khi/\mP^2}\left(K^{\al\bbet}F_\al
F_\bbet-3\frac{\vert W_{\rm
HPS}\vert^2}{\mP^2}\right)\>\>\>\mbox{and}\>\>\>K_{\al\bbet}=\frac{\partial^2
\Khi}{\partial\phi^\al\partial\phi^{*\bbet}},\label{Vsugra}
\end{equation}
with $K^{\bbet\al}K_{\al\bar \gamma}=\delta^\bbet_{\bar \gamma}$ and
$F_\al=W_{\rm HPS,\phi^\al} +K_{,\phi^\al}W_{\rm HPS}/\mP^2$.

If we parameterize the SM neutral components of $\Hcc$ and $\bHc$ by
\beq\label{hpar}
\snH=he^{i\th}\cos\theta_\nu/\sqrt{2}\>\>\>\mbox{and}\>\>\>\snHb=he^{i\thb}\sin\theta_\nu/\sqrt{2},\eeq
we can easily deduce from \Eref{Vd1} that a D-flat direction
occurs at
\beq\label{inftr}
\th=\thb=0,\>\thn={\pi/4}\>\>\>\mbox{and}\>\>\>\eH=\eHb=\uH=\uHb=\dH=\dHb=g^c=\bar
g^c=0.\eeq
Along this direction, $V_{\rm HD}$ in \Eref{Vd1} vanishes and so,
$\Vhi=\what V_{\rm HF}$ takes the form
\beq \label{Vhi}\Vhio
=\mP^4\frac{\ld^2(\xsg^2-4\mpq^2)^2}{16f^2}\>\>\>\mbox{with}\>\>\>f=-{\Ohi\over3}=1+\ck
\xsg^2\>\>\>
\mbox{and}\>\>\>\ck=-\frac{1}{6}+\frac{\kn}{4}\cdot\eeq
Here $\mpq=\Mpq/\mP$ and $\xsg=h/\mP$.  From \Eref{Vhi}, we can
verify that for $\ck\gg1$, $\Shi$ in \Eref{Snpq} takes a form
suitable for the realization of non-MHI: the terms in the
ellipsis vanish, and more importantly $\Vhi$ develops a plateau since $\mpq\ll1$ --
see \Sref{fhi2}. The (almost) constant potential energy density $\Vhio$ and
the corresponding Hubble parameter $\He_{\rm HI0}$ along the
trajectory in \Eref{inftr} are given by
\beq \Vhio=
{\ld^2\sg^4\over16f^2}\simeq{\ld^2\mP^4\over16\ck^2}\>\>\>\mbox{and}\>\>\>
\He_{\rm
HI0}={\Vhio^{1/2}\over\sqrt{3}\mP}\simeq{\ld\mP\over4\sqrt{3}\ck}\,\cdot\label{Vhio}\eeq

We next proceed to check the stability of the trajectory in \Eref{inftr}
w.r.t the fluctuations of the various fields. To this end, we expand them in
real and imaginary parts as follows
\bea X= {x_1+ix_2\over\sqrt{2}},\>\>\>\bar X= {\bar x_1+i\bar
x_2\over\sqrt{2}}\>\>\>\mbox{where}\>\>\>X=\eH,\uH,\dH, g^c
\>\>\>\mbox{and}\>\>\>x=e, u, d, g~. \label{cannor} \eea
respectively. Notice that the field $S$ can be rotated to the real
axis via a suitable R transformation. Since along the trajectory
in \Eref{inftr} the various fields, $X$ and $\bar X$, are confined
to zero, the radial parametrization employed in \Eref{hpar} is not
convenient here. Performing a Weyl transformation as described in
detail in \cref{nmN}, we obtain
$$ S_{\rm HI}=\int d^4 x \sqrt{-\geu}\left(-\frac12
\mP^2\rce+\lin{\partial_\mu\nu_H^{c*}}{\partial_\mu\bar\nu_H^{c*}}
{M_K\over
f^2}\stl{\partial^\mu\snH}{\partial^\mu\snHb}+\frac{1}{2f}\sum_{\chi}\partial_\mu\chi
\partial^\mu \chi-\Vhi\right),$$ \beq
\mbox{where}\>\>\>M_K=\mtt{\kappa}{\bar\kappa}{\bar\kappa}{\kappa},
\>\>\bar\kappa={3\ck^2\xsg^2},\>\>\kappa=f+\bar\kappa
\>\>\>\mbox{and}\>\>\>\Vhi=\what V_{\rm HF}+ {V_{\rm
HD}/f^2}.\label{Sni1} \eeq
In deriving this result, we take into account that $f_{,\chi}\ll
f_{,\sg}$ with $\chi=S, x_1, x_2, \bar x_1$ and $\bar x_2$, and
keep only terms up to quadratic order in the fluctuations $\chi$
and their derivatives. To canonically normalize the fields $\snH$
and $\snHb$, we have to diagonalize the matrix $M_K$. This can be
realized via a similarity transformation involving an orthogonal
matrix $U_K$ as follows:
\beq \label{diagMk} U_K M_K U_K^\tr =\diag\lf\bar
f,f\rg,\>\>\>\mbox{where}\>\>\>\bar
f=f+6\ck^2\xsg^2\>\>\>\mbox{and}\>\>\>
U_K={1\over\sqrt{2}}\mtt{1}{1}{-1}{1}. \eeq
By inserting $\openone=U_K U_K^\tr= U_K^\tr U_K$ on the left
and the right of $M_K$, we bring the second term of the parenthesis in the
RHS of Eq.~(\ref{Sni1}) into the form
\beq {1\over 2f^2}\Bigg(\bar f\lf \partial_\mu h \partial^\mu
h+{1\over2}h^2\partial_\mu \theta_+ \partial^\mu
\theta_+\rg+fh^2\lf{1\over2}\partial_\mu \theta_- \partial^\mu
\theta_-+\partial_\mu \theta_\nu \partial^\mu \theta_\nu\rg\Bigg),
\eeq
along the trajectory in \Eref{inftr}. Here
$\th_{\pm}=\lf\bar\th\pm\th\rg/\sqrt{2}$. Consequently, we can
introduce the EF canonically normalized fields, $\se, \widehat
\theta_+,\widehat \theta_-, \widehat \theta_\nu$ and $\widehat
\chi$, as follows -- cf. \cref{linde1, linde2}
\beq \label{VJe} \frac{d\se}{dh}=J={\sqrt{\bar f}\over
f}=\sqrt{\frac{1}{f}+{6\ck^2\xsg^2\over f^2}},\>\>\>\widehat
\theta_+ ={Jh\theta_+\over\sqrt{2}},\>\>\>\widehat \theta_-
={h\theta_-\over\sqrt{2f}},\>\>\>\widehat \theta_\nu =
\frac{\sg\theta_\nu}{\sqrt{f}}\>\>\>\mbox{and}\>\>\>\widehat \chi
=\frac{\chi}{\sqrt{f}} \cdot\eeq
Taking into account the approximate expressions for $\dot\sg$ --
where the dot denotes derivation w.r.t the cosmic time, $t$ -- $J$
and the slow-roll parameters $\widehat\epsilon, \,\widehat\eta$,
which are displayed in \Sref{fhi2}, we can verify that, during a
stage of slow-roll non-MHI,  $\dot{\widehat \th_+}\simeq J\sg\dot
\th_+/\sqrt{2}$  since $J\sg\simeq\sqrt{6}\mP$,  $\dot{\widehat
\th_-}\simeq \sg\dot \th_-/\sqrt{2f}$ and  $\dot{\widehat
\th_\nu}\simeq \sg\dot \th_\nu/\sqrt{f}$  since
 $h/\sqrt{f}\simeq\mP/\sqrt{\ck}$, and finally  $\dot{\widehat
\chi}\simeq\dot \chi/\sqrt{f}$. For the latter, the quantity $\dot
f/f^{3/2}$, involved in relating $\dot{\widehat \chi}$ to $\dot
\chi$, turns out to be negligibly small, since $\dot f/
f^{3/2}=f_{,\sg}\dot\sg/
f^{3/2}=-{\ld\sqrt{\widehat\epsilon|\widehat\eta|}\mP/2\sqrt{3}\ck}$.
Therefore the action in \Eref{Sni1} takes the form
\beq S_{\rm HI}=\int d^4 x \sqrt{-\geu}\left(-\frac12 \mP^2
\rce+\frac12\geu^{\mu\nu} \sum_{\phi}\partial_\mu \widehat\phi
\partial_\nu \widehat\phi-\Vhi\right),\>\>\>\mbox{with}\>\>\>\Vhi =
\widehat{V}_{\rm HF}+{V_{\rm HD}\over f^2} \label{Sni} \eeq
where $\phi$ stands for $\sg, \th_+, \th_-,\th_\nu, x_1, x_2, \bar
x_1, \bar x_2$ and $S$.

Having defined the canonically normalized scalar fields, we can
proceed in investigating the stability of the inflationary
trajectory of \Eref{inftr}. To this end, we expand $\Vhi$ in
\Eref{Sni} to quadratic order in the fluctuations around the
direction of \Eref{inftr}, as we describe in detail in
\Sref{mscalars}. In \Tref{tab2} we list the eigenvalues of the
masses-squared matrices
$M^2_{\al\bt}=\lf\partial^2\Vhi/\partial\what\chi_\al\partial\what\chi_\beta\rg$
with $\chi_\al=\th_+, \th_-,\th_\nu, x_{1,2}, {\bar x}_{1,2}$ and
$S$ involved in the expansion of $\Vhi$. We arrange our findings
into three groups: the SM singlet sector, $S-\snH-\snHb$, the
sector with the $\uH,\uHb$ and the $\eH,\eHb$ fields which are
related with the broken generators of $\Ggut$ and the sector with
the $\dH, \dHb$ and the $g^c,\bar g^c$ fields.

As we observe from the relevant eigenvalues of the mass-squared
matrices, no instability -- as the one found in \cref{nmN}
-- arises in the spectrum. In particular, it is evident that
$\kx\gtrsim1$ assists us to achieve $m_{\widehat S}^2>0$ -- in
accordance with the results of \cref{linde2}. Moreover, the D-term
contributions to $m^2_{\what \th_\nu}$ and $m^2_{\what u-}$ --
proportional to the gauge coupling constant $g\simeq0.7$ -- ensure the
positivity of these masses squared. Finally the masses that
the scalars $\what d_{1,2}$ acquire from the second and third
term of the RHS of \Eref{Whi} lead to the positivity of
$m_{\what d2}^2$ for $\lH$ of order unity. We have also
numerically verified that the masses of the various scalars remain
greater than the Hubble parameter during the last $50-60$
e-foldings of non-MHI, and so any inflationary perturbations of
the fields other than the inflaton are safely eliminated.

In \Tref{tab2} we also present the masses squared of the gauge
bosons, chiral fermions and gauginos of the model along the
direction of \Eref{inftr}. The mass spectrum is necessary in order
to calculate the one-loop radiative corrections. Let us stress
here that the non-vanishing values of $\snH$ and $\snHb$ trigger
the spontaneous breaking of $\Ggut$ to $G_{\rm SM}$. In particular
we have the following pattern of symmetry breaking
$$ SU(4)_{\rm C}\times SU(2)_{\rm R} \to SU(3)_{\rm C}\times U(1)_{Y}.$$
Therefore, 9 of the 18 generators of $SU(4)_{\rm C}\times
SU(2)_{\rm R}$ are broken, leading to 9 Goldstone bosons which are
``eaten'' by the 9 gauge bosons which become massive. As a
consequence, 36 \emph{degrees of freedom} (d.o.f) of the spectrum
before the spontaneous breaking (18 d.o.f corresponding to 8
complex scalars, $u_{H\ca}^c,\bar u_{H\ca}^c, \eH$ and $\eHb$, and
2 real scalars, $\th$ and $\bar\th $, and 18 d.o.f corresponding
to 9 massless gauge bosons, $A_{\rm C}^9-A_{\rm C}^{14}$, $A_{\rm
R}^1, A_{\rm R}^{2}$ and $A^\perp$) of $\Ggut$ are redistributed
as follows: 9 d.o.f are associated with the real propagating
scalars ($\th_+,x_{1-}$ and $x_{2+}$ with $x=u^\ca$ and $e$)
whereas the residual 9 d.o.f combine together with the $18$ d.o.f
of the initially massless gauge bosons to make massive the
following combinations of them $A_{\rm C}^{\ca\pm}, A_{\rm
R}^{\pm}$ and $A^{\perp}$ -- see \Sref{mgbosons} of Appendix A.

From \Tref{tab2} we can deduce that the numbers of bosonic and
fermionic d.o.f in each sector are equal. Indeed in the
$S-\snH-\snHb$ sector, we obtain 10 bosonic d.o.f and 10 fermionic
d.o.f. Note that we consider $S$ as a complex field and we take
into account the 1 d.o.f of $h$ which is not perturbed in the
expansion of \Eref{Vexp}. Similarly in the $\uH-\uHb$ and
$\eH-\eHb$ [$\dH-\dHb$ and $g^c-\bar g^c$] sector we obtain 32
[24] bosonic d.o.f and an equal number of fermionic d.o.f. Note
also that the spectrum contains a massless fermion which must be
present due to the spontaneous SUSY breaking caused by the
tree-level potential energy density in \Eref{Vhio}.

\renewcommand{\arraystretch}{1.4}
\begin{table}[!t]
\begin{center}
\begin{tabular}{|c||c|l|}\hline
{\sc Fields}&{\sc Masses Squared}&{\sc Eigenstates} \\
\hline\hline
\multicolumn{3}{|c|}{\sc The $S$ -- $\snH$ -- $\snHb$  Sector}\\
\hline
%
2 real scalars &$ m_{\widehat \theta_\nu}^2=\mP^2\xsg^2\lf2\ld^2
(\xsg^2-6)+15g^2f\rg/24f^2$&$\widehat \theta_\nu$\\
&$m_{\widehat\th+}^2=\ld^2\mP^4\xsg^2\lf1+6\ck\rg/12J^2f^3\simeq4\Hhi^2$&$
\widehat\th_{+}$\\
1 complex scalar &$ m_{\widehat S}^2=\ld^2\mP^2\xsg^2{ \lf 12 +
\xsg^2 \bar f\rg\lf6 \kx f-1\rg/6f^2\bar f}$&$\widehat S$\\
2 gauge  bosons&$
m_{\perp}^2=5g^2\mP^2\xsg^2/8f,\>m^2_{||}=0$&$A^{\perp},
A^{\|}$\\[0.5mm]
\hline
4 Weyl fermions&$ m_{\what\psi_{S\nu}}^2=\ld^2\mP^2\xsg^2/2\bar f
f^2$&
$\widehat \psi_{S\nu\pm}$\\
&$m_{\perp}^2=5g^2\mP^2\xsg^2/8f$&$\ldu^{\perp},
\widehat\psi_{\nu-}$\\
1 Majorana fermion &$m^2_{||}=0$&
$\ldu^{||},\widehat\psi_{\nu+}$\\[0.5mm]
\hline\hline
\multicolumn{3}{|c|}{\sc The $u_{H\ca}^c$ --
$\bar u_{H\ca}^c$ (${\rm a}=1,2,3$) and $\eH$ -- $\eHb$ Sectors}\\
\hline
$2(3+1)$ real scalars &$ m_{\widehat u-}^2=\mP^2\xsg^2\lf\ld^2
(\xsg^2-3)+3g^2f\rg/12f^2$&$\widehat u^{\rm a}_{1-},\>\widehat
u^{\rm a}_{2+},$\\
&$m^2_{\widehat e-}=m^2_{\widehat u-}$&$\widehat e_{1-},\>\widehat e_{2+}$\\
$2(3+1)$ gauge  bosons&$
m_{\pm}^2=g^2\mP^2\xsg^2/4f$&$A_{\rm C}^{\rm a\pm}, A_{\rm R}^{\pm}$\\
\hline
$4(3+1)$ Weyl fermions&$m_{\pm}^2=g^2\mP^2\xsg^2/4f$&
$\ldu_{\rm C}^{\rm a\pm}, \psi^\ca_{u}, \psi^\ca_{\bar u}$\\
&& $\ldu_{\rm R}^{\pm}, \psi_e, \psi_{\bar e}$\\
\hline\hline
\multicolumn{3}{|c|}{\sc The $d_{H\ca}^c$ --
$\bar d_{H\ca}^c$ and $g_\ca^c$ -- $\bar g_\ca^c$ (${\ca}=1,2,3$) Sectors}\\
\hline
$3\cdot 8$ real scalars &$ m_{\widehat g}^2=\mP^2\xsg^2\lf\ld^2
\xsg^2+24\lHb^2f\rg/24f^2$&$\widehat g^{\rm a}_1,\widehat g^{\rm a}_2$\\
&$ m_{\widehat{\bar g}}^2=\mP^2\xsg^2\lf\ld^2
\xsg^2+24\lH^2f\rg/24f^2$&$\widehat{\bar g}^{\rm a}_1,\widehat{\bar g}^{\rm a}_2$\\
&$ m_{\widehat d+}^2=\mP^2\xsg^2\lf\ld^2
+4\lH^2f\rg/4f^2$&$\widehat{d}^{\rm a}_{1+},\widehat{d}^{\rm a}_{2-}$\\
&$ m_{\widehat d-}^2=\mP^2\xsg^2\lf\ld^2
\lf\xsg^2-3\rg+12\lH^2f\rg/12f^2$&$\widehat{d}^{\rm
a}_{1-},\widehat{d}^{\rm
a}_{2+}$\\[1mm]
\hline
$3\cdot4$ Weyl fermions& $m_{\widehat\psi_{\bar g d}}^2=\lH^2\mP^2\xsg^2/f$
&$\widehat\psi^{\rm a}_{\bar g d\pm}$\\
&$m_{\widehat\psi_{g\bar
d}}^2=\lHb^2\mP^2\xsg^2/f$&$\widehat\psi^{\rm a}_{g\bar
d\pm}$\\[1.5mm]
\hline
\end{tabular}\end{center}
\vchcaption{\sl\ftn The mass spectrum of our model along the
inflationary trajectory of \Eref{inftr}. To avoid very lengthy
formulas we neglect terms proportional to $\mpq^2$ and we assume
$\lH\simeq\lHb$ for the derivation of the masses of the scalars in
the superfields $\dH$ and $\dHb$. The various eigenstates are
defined in \Sref{fhi1} and Appendix A.}\label{tab2}
\end{table}

The $8$ Goldstone bosons, associated with the modes $x_{1+}$ and
$x_{2-}$ with $x=u^\ca$ and $e$, are not exactly massless since
$\what V_{{\rm HI},h}\neq0$ -- contrary to the situation of
\cref{jean} where the direction with non vanishing $\vev{\snH}$
corresponds to a minimum of the potential. These masses turn out
to be $m_{x0}=\ld\mP\xsg/2f$. On the contrary, the angular
parametrization in \Eref{hpar} assists us to isolate the massless
mode $\what\th_-$, in agreement with the analysis of
\cref{linde1}. In computing below the one-loop radiative
corrections, $V_{\rm rc}$, in our model, we do not treat the
residual Goldstone bosons as independent fields, since their
associated d.o.f are ``eaten'' by the massive gauge bosons -- for
a different point of view, see \cref{postma}. After all, as we
expect and verified numerically, $V_{\rm rc}$ has no significant
effect on the inflationary dynamics and predictions, since the
slope of the inflationary path is generated at the classical level
-- see the expressions for $\widehat\epsilon$ and $\widehat\eta$
below -- and no significant running of the relevant parameters
occurs -- contrary to the SM or next-to-MSSM non-MHI. Employing
the well-known Coleman-Weinberg formula \cite{cw}, we find
\beq V_{\rm rc}= V_{S\snH\snHb}+V_{\uH\uHb\eH\eHb}+V_{\dH\dHb g^c
\bar g^c},\eeq
where the individual contributions, coming from the corresponding
sectors of Table~\ref{tab2}, are given by
\beqs\baq  V_{S\snH\snHb}&=& {1\over64\pi^2}\lf m_{\widehat
\th_\nu}^4\ln{m_{\widehat \th_\nu}^2\over\Lambda^2} + m_{\widehat
\th_+}^4\ln{m_{\widehat \th_+}^2\over\Lambda^2} +2 m_{\widehat
S}^4\ln{m_{\widehat S}^2\over\Lambda^2}
\nonumber \right.\\
&+&\left.3m_{\perp}^4\ln{m_{\perp}^2\over\Lambda^2}
-4m_{\widehat\psi_{S\nu+}}^4\ln{m_{\widehat\psi_{S\nu+}}^2\over\Lambda^2}
-4m_{\perp}^4\ln{m_{\perp}^2\over\Lambda^2}\rg,\\
V_{\uH\uHb\eH\eHb}&=&{4\over64\pi^2}\lf2m_{\widehat
u-}^4\ln{m_{\widehat
u-}^2\over\Lambda^2}+6m_{\pm}^4\ln{m_{\pm}^2\over\Lambda^2}-8m_{\pm}^4\ln{m_{\pm}^2\over\Lambda^2}\rg,\\
V_{\dH\dHb g^c \bar g^c}&=& {3\over64\pi^2}\lf 2m_{\widehat
g}^4\ln{m_{\widehat g}^2\over\Lambda^2} + 2m_{\widehat{\bar
g}}^4\ln{m_{\widehat{\bar g}}^2\over\Lambda^2}+2m_{\widehat
d+}^4\ln{m_{\widehat d+}^2\over\Lambda^2}+2m_{\widehat
d-}^4\ln{m_{\widehat d-}^2\over\Lambda^2}\right.\nonumber\\ &-&
\left.4m_{\widehat{\psi}_{\bar gd}}^4\ln{m_{\widehat{\psi}_{\bar
gd}}^2\over\Lambda^2} - 4m_{\widehat{\psi}_{g\bar
d}}^4\ln{m_{\widehat{\psi}_{g\bar d}}^2\over\Lambda^2}\rg~.
\label{Vrcs}  \eaq\eeqs
Here $\Lambda$ is a renormalization mass scale. Based on the
action of \Eref{Sni} with $\Vhi\simeq\Vhio+V_{\rm rc}$, we can
proceed to the analysis of \FHI in the EF, employing the standard
slow-roll approximation \cite{review, lectures}. It can be shown
\cite{general} that the results calculated this way are the same
as if we had calculated them using the non-minimally coupled
scalar field in the JF.

\subsection{The Inflationary Observables -- Requirements}\label{fhi2}

Under the assumption that there is a conventional cosmological
evolution (see below) after non-MHI, the model parameters can be
restricted, imposing the following requirements:

\paragraph{3.2.1} According to the inflationary paradigm,
the horizon and flatness problems of the standard Big Bag
cosmology can be successfully resolved provided that the number of
e-foldings, $\widehat N_{*}$, that the scale $k_{*}=0.002/{\rm
Mpc}$ suffers during \FHI takes a certain value, which depends on
the details of the cosmological model. The required $\widehat
N_{*}$ can be easily derived \cite{hinova}, consistently with our
assumption of a conventional post-inflationary evolution. In
particular, we assume that non-MHI is followed successively by the
following three epochs: {\sf (i)} the decaying-inflaton dominated
era which lasts until the reheating temperature $\Trh$, {\sf (ii)} a
radiation dominated epoch, with initial temperature $\Trh$, which
terminates at the matter-radiation equality, {\sf (iii)} the
matter dominated era until today. Employing standard methods
\cite{nmi, hinova}, we can easily derive the required
$\widehat{N}_{*}$ for our model, with the result:
\begin{equation}  \label{Ntot}
\widehat{N}_{*}\simeq22.5+2\ln{V_{\rm
HI}(\sg_{*})^{1/4}\over{1~{\rm GeV}}}-{4\over 3}\ln{V_{\rm
HI}(\sg_{\rm f})^{1/4}\over{1~{\rm GeV}}}+ {1\over3}\ln {T_{\rm
rh}\over{1~{\rm GeV}}}+{1\over2}\ln{f(\sg_{\rm f})\over
f(\sg_*)}\cdot
\end{equation}
On the other hand, $\widehat N_{*}$ can be calculated via the
relation
\begin{equation}
\label{Nhi}  \widehat{N}_{*}=\:\frac{1}{m^2_{\rm P}}\;
\int_{\se_{\rm f}}^{\se_{*}}\, d\se\: \frac{\Vhi}{\Ve_{{\rm
HI},\se}}= {1\over\mP^2}\int_{h_{\rm f}}^{h_{*}}\, d\sg\:
J^2\frac{\Ve_{\rm HI}}{\Ve_{{\rm HI},h}},
\end{equation}
where $h_*~[\se_*]$ is the value of $\sg~[\se]$ when $k_*$ crosses
the inflationary horizon. Also $h_{\rm f}~[\se_{\rm f}]$ is the
value of $\sg~[\se]$ at the end of \FHI determined, in the
slow-roll approximation, by the condition -- see e.g.
\cref{review, lectures}:
$$ {\ftn\sf
max}\{\widehat\epsilon(\sgf),|\widehat\eta(\sgf)|\}=1,\>\>\>\mbox{where}$$
\beqs\baq &&  \label{sr1}\widehat\epsilon=
{\mP^2\over2}\left(\frac{\Ve_{{\rm HI},\se}}{\Ve_{\rm
HI}}\right)^2={\mP^2\over2J^2}\left(\frac{\Ve_{{\rm
HI},h}}{\Ve_{\rm HI}}
\right)^2\simeq {4\mP^4\lf1+4\ck\mpq^2\rg^2\over3\ck^2\sg^4}\\
\mbox{and}\>\>\> && \label{sr2}\widehat\eta= m^2_{\rm
P}~\frac{\Ve_{{\rm HI},\se\se}}{\Ve_{\rm HI}}={\mP^2\over
J^2}\left( \frac{\Ve_{{\rm HI},hh}}{\Ve_{\rm HI}}-\frac{\Ve_{{\rm
HI},h}}{\Ve_{\rm HI}}{J_{,\sg}\over
J}\right)\simeq-{4\mP^2\lf1+4\ck\mpq^2\rg\over3\ck\sg^2}~,
\eaq\eeqs
are the slow-roll parameters. Here we employ \Eref{Vhio} and the
following approximate relations:
\beq \label{help} J\simeq \sqrt{6}{\mP\over\sg},\>\>\>\widehat
V_{{\rm HI},\sg}\simeq {4\Vhi\over \ck h^3}\mP^2\lf1+4\ck\mpq^2\rg
\>\>\>\mbox{and}\>\>\>\widehat V_{{\rm
HI},\sg\sg}\simeq-{12\Vhi\over \ck
h^4}\mP^2\lf1+4\ck\mpq^2\rg.\eeq
The numerical computation reveals that \FHI terminates due to the
violation of the $\widehat\epsilon$ criterion at a value of $\sg$
equal to $\sgf$, which is calculated to be
\beq \widehat\epsilon\lf\sgf\rg=1\>\Rightarrow\>
\sgf=\lf{4/3}\rg^{1/4}\mP\sqrt{\lf1+4\ck\mpq^2\rg/\ck}.
\label{sgap}\eeq
Given that $\sgf\ll\sg_*$, we can write $\sg_*$ as a function of
$\widehat{N}_{*}$ as follows
\beq \label{s*}
\widehat{N}_{*}\simeq{3\ck\over4}{\sg_*^2-\sgf^2\over
(1+4\ck\mpq^2)\mP^2}\>\Rightarrow\>\sg_*=2\mP\sqrt{\widehat
N_*\lf1+4\ck\mpq^2\rg/3\ck}\cdot\eeq

\paragraph{3.2.2} The power spectrum $P_{\cal R}$ of the curvature
perturbations generated by $h$ at the pivot scale $k_{*}$ is to be
confronted with the WMAP7 data~\cite{wmap}, i.e.
\begin{equation}  \label{Prob}
P^{1/2}_{\cal R}=\: \frac{1}{2\sqrt{3}\, \pi\mP^3} \;
\frac{\Ve_{\rm HI}(\sex)^{3/2}}{|\Ve_{{\rm HI},\se}(\sex)|}=
{1\over2\pi\mP^2}\,\sqrt{\frac{\Vhi(\sg_*)}{6\, \widehat\epsilon\,
(\sg_*)}} \simeq4.93\cdot 10^{-5}. \eeq
Note that since the scalars listed in \Tref{tab2} are massive
enough during non-MHI, the curvature perturbations generated by
$\sg$ are solely responsible for $P_{\cal R}$. Substituting
\eqs{sr1}{s*} into the relation above, we obtain
\beq P^{1/2}_{\cal
R}\simeq{\ld\sg_*^2\over16\sqrt{2}\pi\mP^2\lf1+4\ck\mpq^2\rg}\simeq
{\ld\widehat{N}_{*}\over12\sqrt{2}\pi\ck}\cdot\eeq
Combining the last equality with \Eref{Prob}, we find that
$\ld$ is to be proportional to $\ck$, for almost constant $\Ne_*$.
Indeed, we obtain
\beq \ld\simeq{8.4\cdot10^{-4}\pi\ck/\Ne_*}\>\Rightarrow\>
\ck\simeq20925\ld\>\>\>\mbox{for}\>\>\>\Ne_*\simeq55.\label{lan}\eeq

\paragraph{3.2.3} The (scalar) spectral index $n_{\rm s}$, its
running $a_{\rm s}$, and the scalar-to-tensor ratio $r$ must be
consistent with the fitting \cite{wmap} of the WMAP7, BAO and
$H_0$ data, i.e.,
\begin{equation}  \label{obs}
\mbox{\ftn\sf (a)}\>\>\>\ns=0.968\pm0.024,\>\>\>\mbox{\ftn\sf
(b)}\>\>-0.062\leq a_{\rm s}\leq0.018
\>\>\>\mbox{and}\>\>\>\mbox{\ftn\sf (c)}\>\>r<0.24
\end{equation}
at 95$\%$ \emph{confidence level} (c.l.). The observable
quantities above can be estimated through the relations:
\beqs\baq \label{ns} && n_{\rm s}=\: 1-6\widehat\epsilon_*\ +\
2\widehat\eta_*\simeq1-{2/\widehat N_*}, \>\>\> \\
&& \label{as} \alpha_{\rm s}
=\:{2\over3}\left(4\widehat\eta_*^2-(n_{\rm
s}-1)^2\right)-2\widehat\xi_*\simeq-2\widehat\xi_*\simeq{-2/\widehat N^2_*}\>\>\> \\
\mbox{and}\>\>\> && \label{rt}
r=16\widehat\epsilon_*\simeq{12/\widehat N^2_*}, \eaq\eeqs
where $\widehat\xi=\mP^4 {\Ve_{{\rm HI},\sg} \Ve_{{\rm
HI},\se\se\se}/\Vhi^2}=
\mP\,\sqrt{2\widehat\epsilon}\,\widehat\eta_{,h}/
J+2\widehat\eta\widehat\epsilon$. The variables with subscript $*$
are evaluated at $h=h_{*}$ and \eqs{sr1}{sr2} have been employed.

\paragraph{3.2.4} The scale $\Mpq$ can be determined by requiring that
the v.e.vs of the Higgs fields take the values dictated by the
unification of the gauge couplings within the MSSM. Since the
highest mass scale of the model -- see \Tref{tab2} -- in the SUSY
vacuum, \Eref{vevs}, is
\beq \label{mvev} m_{\perp0} = \sqrt{5/2f_0}
g|\vev{\snH}|\>\>\>\mbox{with}\>\>\>f_0=f\lf\vev{h}\rg=1+4\ck\mpq^2
\eeq
(recall that $\mpq=\Mpq/\mP$) we can identify it with the
unification scale $\Mgut$, i.e.
\beq \label{Mpqf} m_{\perp}=\sqrt{5\over2}{g\Mpq\over
f_0}=\Mgut\>\Rightarrow\>\Mpq={\sqrt{2}\Mgut\mP\over
(5g^2\mP^2-2\ck\Mgut^2)^{1/2}}\eeq
The requirement $5g^2\mP^2>2\ck\Mgut^2$ sets an upper bound on
$\ck$, which however can be significantly lowered if we impose the
requirement of Sec. 3.2.1 -- see below. When $\ck$ ranges within
its allowed region, we take
$\Mpq\simeq\lf1.81-2.2\rg\cdot10^{16}~\GeV$.

\paragraph{3.2.5} For the realization of \FHI, we assume that $\ck$ takes relatively
large values -- see e.g. \Eref{Sni1}. This assumption may
\cite{cutoff, unitarizing} jeopardize the validity of the
classical approximation, on which the analysis of the inflationary
behavior is based. To avoid this inconsistency -- which is rather
questionable \cite{cutoff, linde2} though -- we have to check the
hierarchy between the ultraviolet cut-off, $\Ld=\mP/\ck$, of the
effective theory and the inflationary scale, which is represented
by $\Vhi(\sg_*)^{1/4}$ or, less restrictively, by the
corresponding Hubble parameter, $\widehat
H_*=\Vhi(\sg_*)^{1/2}/\sqrt{3}\mP$. In particular, the validity of
the effective theory implies \cite{cutoff}
\beq \label{Vl}\mbox{\ftn\sf (a)}\>\>\> \Vhi(\sg_*)^{1/4}\leq\Ld
\>\>\>\mbox{or}\>\>\>\mbox{\ftn\sf (b)}\>\>\> \widehat
H_*\leq\Ld\>\>\>\mbox{for}\>\>\>\mbox{\ftn\sf
(c)}\>\>\>\ck\geq1.\eeq

\subsection{Numerical Results}\label{num1}

As can be easily seen from the relevant expressions above, the
inflationary dynamics of our model depends on the following
parameters:
$$\ld,\>\lH,\>\lHb,\>\kx,\>\ck\>\>\>\mbox{and}\>\>\>\Trh.$$
Recall that we determine $\Mpq$ via \Eref{Mpqf} with $g=0.7$. Our
results are essentially independent of $\lH,\>\lHb$ and $\kx$,
provided that we choose some relatively large values for these so
as $m^2_{\what u-}, m^2_{\what d-}$ and $m_{\what S}^2$ in
\Tref{tab2} are positive for $\ld<1$. We therefore set
$\lH=\lHb=0.5$ and $\kx=1$ throughout our calculation. Finally
$\Trh$ can be calculated self-consistently in our model as a
function of the inflaton mass, $\msn$ and the mass $\mrh[{\rm I}]$
of the RH neutrino into which inflaton decays, and the unified
Yukawa coupling constant  $y_{33}$ -- see \Sref{lept}. However the
inflationary predictions depend very weakly on $\Trh$, because
$\Trh$ appears in Eq.~(\ref{Ntot}) through the one third of its
logarithm, and consequently its variation upon some orders of
magnitude has a minor impact on the required value of $\Ne_*$,
which remains almost constant and close to $55$.

\begin{figure}[!t]\vspace*{-.15in}
\hspace*{-.19in}
\begin{minipage}{8in}
\epsfig{file=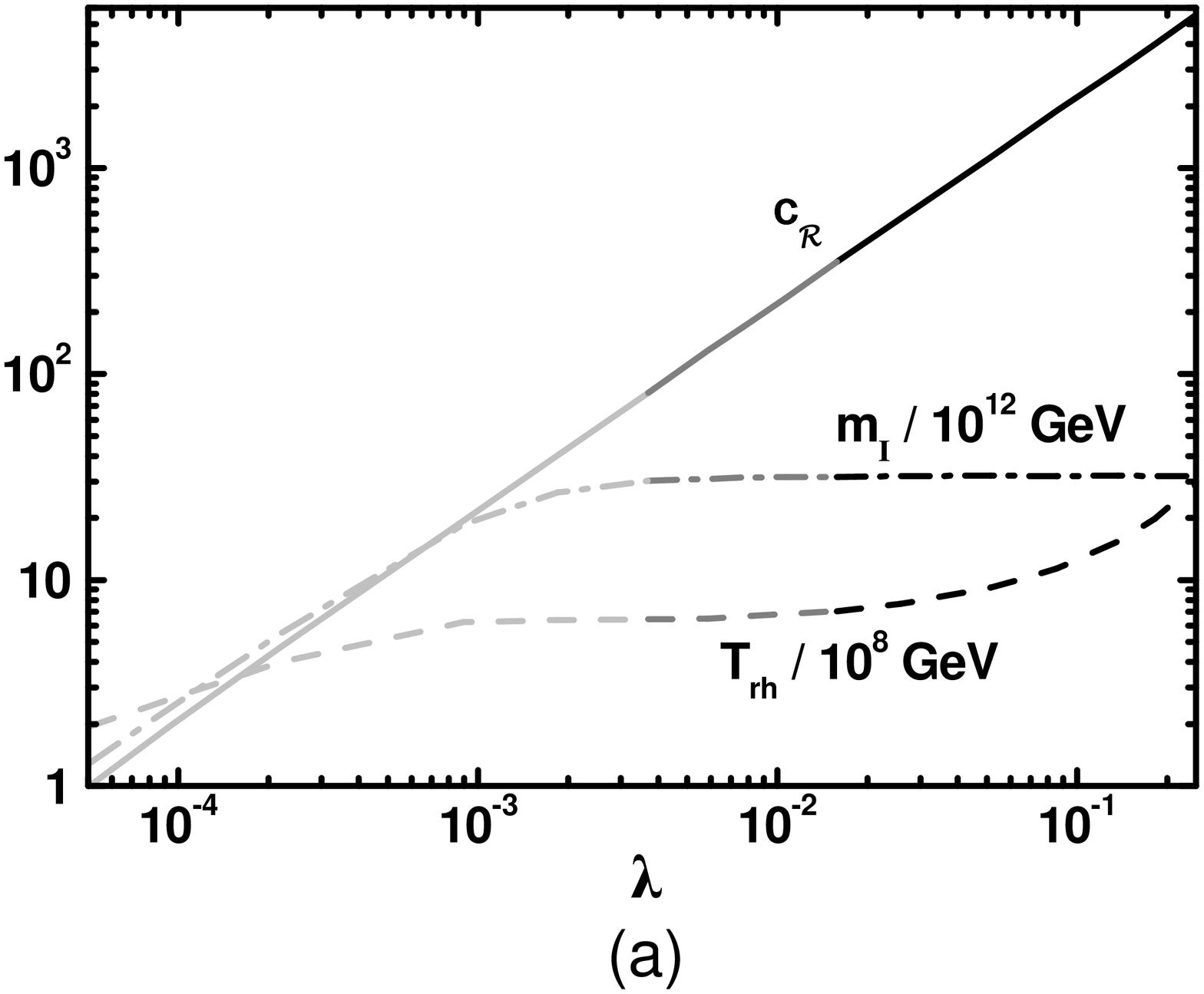,height=3.6in,angle=-90}
\hspace*{-1.2cm}
\epsfig{file=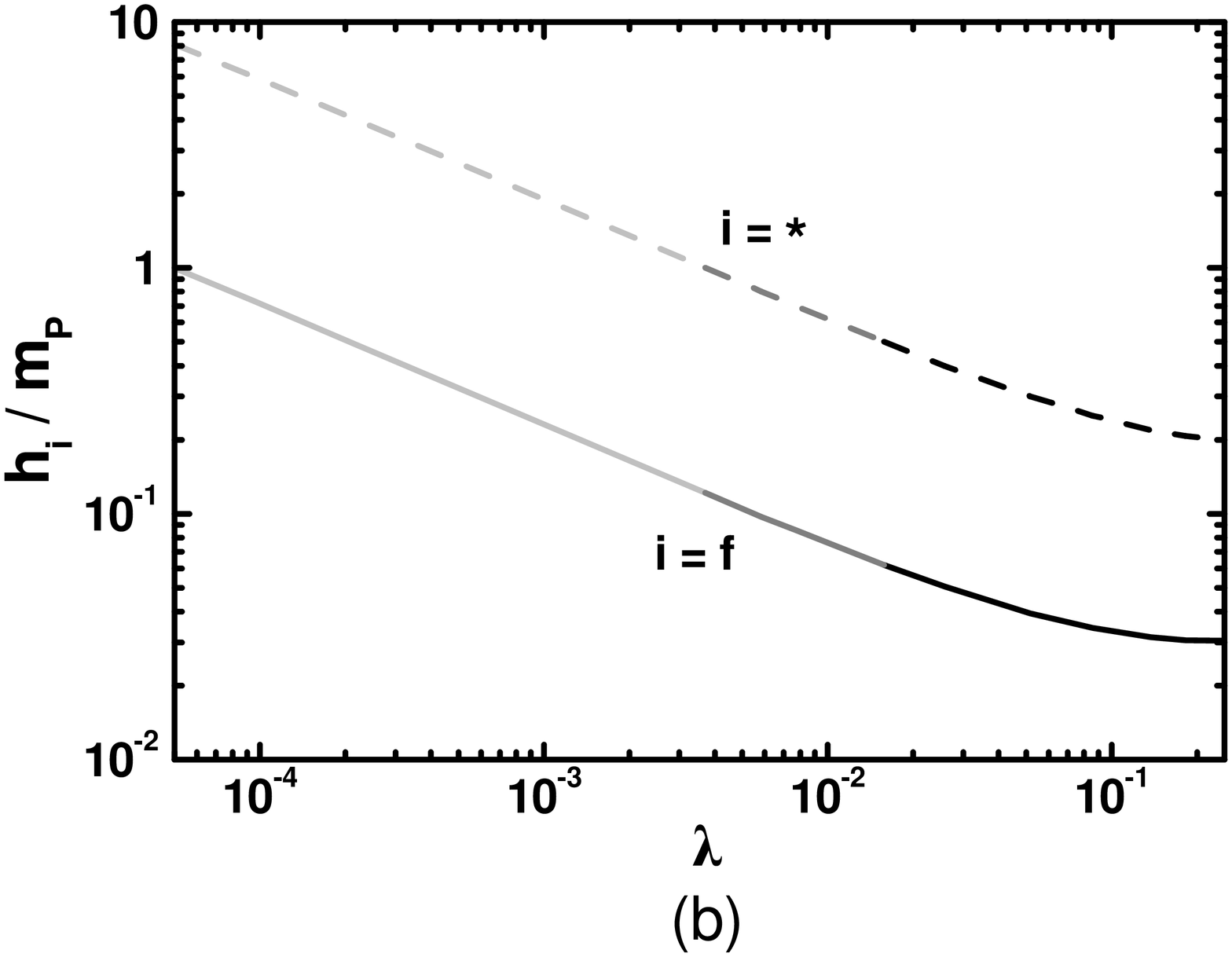,height=3.6in,angle=-90} \hfill
\end{minipage}
\hfill \vchcaption[]{\sl\small  The allowed by Eqs.~(\ref{Ntot}),
(\ref{Prob}), (\ref{Vl}{\sf\ftn b}) and (\ref{Vl}{\sf\ftn c})
values of $\ck$ (solid line), $\Trh$ -- given by \Eref{T1rh} --
(dashed line) and $\msn$ (dot-dashed line) [$\sg_{\rm f}$ (solid
line) and $\sg_*$ (dashed line)] versus $\ld$ (a) [(b)] for
$\kx=1,~\lH=\lHb=0.5$, $\mrh[{\rm I}]=10^{11}~\GeV$ and
$y_{33}=0.5$. The light gray and gray segments denote values of
the various quantities satisfying \sEref{Vl}{a} too, whereas along
the light gray segments we obtain $\sg_*\geq\mP$.}\label{fig1}
\end{figure}


In our numerical code, we use as input parameters $\sg_*,\mrh[{\rm
I}]$ and $\ck$. For every chosen $\ck\geq1$ and $\mrh[{\rm I}]$,
we restrict $\ld$ and $\sg_*$ so that the conditions \Eref{Ntot}
and (\ref{Prob}) are satisfied. In our numerical calculations, we
use the complete formulas for the slow-roll parameters and
$P_{\cal R}^{1/2}$ in \eqss{sr1}{sr2}{Prob} and not the
approximate relations listed in \Sref{fhi2} for the sake of
presentation. Our results are displayed in \Fref{fig1}, where we
draw the allowed values of $\ck$ (solid line), $\Trh$ (dashed
line) and the inflaton mass at the SUSY vacuum -- see \Sref{lept}
--  $\msn$ (dot-dashed line) [$\sg_{\rm f}$ (solid line) and
$\sg_*$ (dashed line)] versus $\ld$ (a) [(b)] for $\mrh[{\rm
I}]=10^{11}~\GeV$. Note that the decay of the inflaton into a RH
neutrino with the mass above is kinematically permitted, for the
depicted $\ld$'s. The lower bound of the depicted lines comes from
the saturation of the \sEref{Vl}{c}. The constraint of
\sEref{Vl}{b} is satisfied along the various curves whereas
\sEref{Vl}{a} is valid only along the gray and light gray segments
of these. Along the light gray segments, though, we obtain
$\sg_*\geq\mP$. The latter regions of parameter space are not
necessarily excluded \cite{circ}, since the energy density of the
inflaton remains sub-Planckian and so, corrections from quantum
gravity can still be assumed to be small. As $\ck$ increases
beyond $5.6\cdot10^3$, $4\ck\mpq$ becomes larger than $1$, $\what
N_*$ derived by \Eref{s*} starts decreasing and therefore, non-MHI
fails to fulfil the relevant requirement. All in all, we obtain
\beq\label{res1} 1\lesssim
\ck\lesssim5.6\cdot10^3\>\>\>\mbox{and}\>\>\>5\cdot10^{-5}\lesssim
\ld\lesssim0.25\>\>\>\mbox{for}\>\>\> 53.9\lesssim
\Ne_*\lesssim55.\eeq

From \sFref{fig1}{a}, we can verify our analytical estimation in
\Eref{lan} according to which $\ld$ is proportional to $\ck$. On the
other hand, the variation of $\sg_{\rm f}$ and $\sg_*$ as a function
of $\ck$ -- drawn in \sFref{fig1}{b} -- is consistent with
\eqs{sgap}{s*}. Note that the inclusion of the term $4\ck\Mpq^2$ in
the numerators of these relations is crucial in order to obtain a
reliable result for $\ld\gtrsim0.13$ or $\ck\gtrsim3\cdot10^3$ --
cf. \cref{nmN}. Letting $\ld$ or $\ck$ vary within its allowed
region in \Eref{res1}, we obtain
\beq\label{res} 0.964\lesssim \ns\lesssim0.965,\>\>\>-6.5\lesssim
{\as\over10^{-4}}\lesssim-6.2\>\>\>\mbox{and}\>\>\>4.2\gtrsim
{r\over10^{-3}}\gtrsim3.5.\eeq
Clearly, the predicted $\as$ and $r$ lie within the allowed ranges
given in \sEref{obs}{b} and \sEref{obs}{c} respectively, whereas
$\ns$ turns out to be impressively close to its central
observationally favored value -- see \sEref{obs}{a} and cf.
\cref{linde2}.

\section{Non-Thermal Leptogenesis}\label{pfhi}

In this section, we specify how the SUSY inflationary scenario
makes a transition to the radiation dominated era, and give an
explanation of the origin of the observed BAU consistently with
the $\Gr$ constraint. The main features of the post-inflationary
evolution of our model are described in \Sref{lept}. In
\Sref{cont1} we describe the additional constraints that we impose
on our setting, and finally we delineate the allowed parameter
space of our cosmological model in Sec.~\ref{num}.

\subsection{The General Set-up}\label{lept}

When \FHI is over, the inflaton continues to roll down towards the
SUSY vacuum, \Eref{vevs}. There is a brief stage of tachyonic
preheating \cite{preheating} which does not lead to significant
particle production \cite{garcia}.  Soon after, as discussed in the
Appendix B, the inflaton settles into a phase of damped oscillations
initially around zero -- where $\Vhio$ has a maximum -- and then
around one of the minima of $\Vhio$. Whenever the inflaton passes
through zero, particle production may occur creating mostly
superheavy bosons via the mechanism of instant preheating
\cite{instant}. This process becomes more efficient as $\ld$
decreases, and further numerical investigation is required in order
to check the viability of our leptogenesis scenario detailed below.
For this reason, we restrict to $\ld$'s larger than $0.001$, which
ensures a less frequent passage of the inflaton through zero,
weakening thereby the effects from instant preheating and other
parametric resonance effects -- see Appendix B. Intuitively the
reason is that larger $\ld$'s require larger $\ck$'s, see
\Eref{lan}, diminishing therefore $h_{\rm f}$ given by \Eref{s*},
which sets the amplitude of the very first oscillations.

\renewcommand{\arraystretch}{1.3}
\begin{table}[!t]
\begin{center}
\begin{tabular}{|c|c||c|c|}\hline
{\sc Eigenstates}&{\sc Masses }&{\sc Eigenstates}&{\sc Masses} \\
\hline\hline
\multicolumn{2}{|c||}{\sc The $S$ -- $\snH$ -- $\snHb$
Sector}&\multicolumn{2}{|c|}{\sc The $u_{H\ca}^c$ -- $\bar
u_{H\ca}^c$ and $\eH$ -- $\eHb$ Sectors}\\ \hline
$\what{\delta h}$& $\msn=\sqrt{2}\ld\Mpq/\vev{J}f_0$ & $\widehat
u^{\rm a}_{1-},\>\widehat
u^{\rm a}_{2+},\widehat e_{1-},\>\widehat e_{2+}$&$g\Mpq/\sqrt{f_0}$\\
$\what\theta_\nu$& $\sqrt{5/2f_0}g\Mpq$& $A_{\rm C}^{\rm a\pm},
A_{\rm R}^{\pm}$&$g\Mpq/\sqrt{f_0}$\\\cline{3-4}
$\what\theta_+$& $\sqrt{2}\ld\Mpq/J_0f_0$ & $\ldu_{\rm C}^{\rm
a\pm},
\psi^\ca_{u}, \psi^\ca_{\bar u}$&$g\Mpq/\sqrt{f_0}$\\
$\what S$& $\sqrt{2}\ld\Mpq/\sqrt{\bar f_0}$& $\ldu_{\rm R}^{\pm},
\psi_e, \psi_{\bar e}$&$g\Mpq/\sqrt{f_0}$\\\cline{3-4}
$A^{\perp}, A^{\|}$&$\sqrt{5/2f_0}g\Mpq$ &\multicolumn{2}{|c|}{\sc
The $d_{H\ca}^c$ -- $\bar d_{H\ca}^c$ and $g^c_\ca$ -- $\bar
g^c_\ca$ Sectors}\\\cline{3-4}
$A^{\|}$&$0$&$\widehat g^{\rm a}_1,\widehat g^{\rm
a}_2$&$2\lH\Mpq/\sqrt{f_0}$
\\\cline{1-2}
$\widehat \psi_{S\nu\pm}$&$\sqrt{2}\ld\Mpq /\sqrt{\bar f_0}$&
$\widehat{\bar g}^{\rm a}_1,\widehat{\bar g}^{\rm a}_2$& $2\lHb\Mpq/\sqrt{f_0}$\\
$\ldu^{\perp}, \widehat\psi_{\nu-}$& $\sqrt{5/2f_0}g\Mpq$&
$\widehat{d}^{\rm a}_{1},\widehat{d}^{\rm a}_{2}$& $2\lH\Mpq/\sqrt{f_0}$\\
$\ldu^{||},\widehat\psi_{\nu+}$& $0$&$\widehat{\bar d}^{\rm
a}_{1},\widehat{\bar d}^{\rm a}_{2}$& $2\lHb\Mpq/\sqrt{f_0}$\\
\cline{1-2} \cline{3-4}
\multicolumn{2}{|c||}{\sc The $\nu_i^c$ Sector}&$\widehat\psi^{\rm
a}_{\bar g d\pm}$&$2\lH\Mpq/\sqrt{f_0}$\\ \cline{1-2}
$\what\nu_i^c, \what{\tilde\nu}_i^c$&
$\mrh[i]=2\ld_{i\nu^c}\Mpq^2/\Ms\sqrt{f_0}$ & $\widehat\psi^{\rm
a}_{g\bar d\pm}$&$2\lHb\Mpq/\sqrt{f_0}$\\
[1.5mm] \hline
\end{tabular}\end{center}
\vchcaption{\sl\ftn The mass spectrum of our model at the SUSY
vacuum of \Eref{vevs}. We use the abbreviations $\vev{J}=J\lf
h=2\Mpq\rg, f_0=1+4\ck\mpq^2$ and $\bar f_0=
f_0+24\ck^2\mpq^2\simeq J_0^2$. The various eigenstates and
symbols are specified in \Sref{fhi1} and Appendix A.}\label{tab3}
\end{table}

Nonetheless the standard perturbative approach to the inflaton
decay provides a very efficient decay rate. This is to be
contrasted with the SM (or next-to-MSSM) non-MHI, where the
consideration of non-perturbative effects is imperative
\cite{garcia} in order to obtain successful reheating. Namely, at
the SUSY vacuum $\snH$ and $\snHb$ acquire the v.e.vs shown in
\Eref{vevs} giving rise to the mass spectrum presented in
\Tref{tab3}. Note that the masses of the various scalars --
contrary to the masses of the fermions and gauge bosons -- are not
derived from the corresponding formulas listed in \Tref{tab2} with
the naive replacement $\xsg=2\mpq$, since terms proportional to
$\mpq^2$ are neglected there. In \Tref{tab3} we also show the
mass, $\msn$, of the (canonically normalized) inflaton
$\what{\delta h}=\lf h-2\Mpq\rg/J_0$ and the masses $\mrh[i]$ of
the RH neutrinos, $\nu_i^c$, which play a crucial role in our
leptogenesis scenario -- we assume the existence of a term similar
to the second one inside $\ln$ of \Eref{Kcom} for $\nu_i^c$ too.
From \Fref{fig1} we notice that $\msn$ increases with $\ld$ -- as
in the case of HI, cf. \cref{jean} -- only for $\ld\lesssim0.0013$
or $\ck\leq30$. For larger $\ld$'s $\vev{J}=J(h=2\Mpq)$ ranges
from $3$ to $90$ and so $\msn$ is kept independent of $\ld$ and
almost constant at the level of $10^{13}~\GeV$. Indeed, if we
express $\what{\delta h}$ as a function of ${\delta h}$ through
the relation
\beq \label{Jo} {\what{\delta h}\over{\delta h}}\simeq
J_0\>\>\>\mbox{where}\>\>\>
J_0=\sqrt{1+{3\over2}\mP^2f^2_{,\sg}\lf\vev{h}\rg}=\sqrt{1+24\ck^2\mpq^2}\eeq
is obtained by expanding $J$ given in \Eref{VJe} at leading order
in $\xsg$, we find
\beq \label{mqa} \msn\simeq{\sqrt{2}\ld\Mpq\over
f_0J_0}\simeq{\ld\mP\over2\sqrt{3}\ck}\simeq{10^{-4}\mP\over4.2\sqrt{3}}\simeq3\cdot10^{13}~\GeV
\>\>\>\mbox{for}\>\>\>\ld\gtrsim{10^{-4}\over4.2\sqrt{6}\mpq}\simeq1.3\cdot10^{-3}\eeq
where we make use of \Eref{lan} -- note that $f_0\simeq1$. Apart
from some fields in the $\nu_i^c$ sector, $\what{\delta h}$ is the
lightest among the massive particles listed in \Tref{tab3} for
$\ld$ given in \Eref{res1}, since $\vev{J}\gg1$ and $g,\lH$ and
$\lHb$ are taken larger than $\ld$. Therefore perturbative decays
of $\what{\delta h}$ into these massive particles are
kinematically forbidden. For the same reason narrow parametric
resonance \cite{preheating} effects are absent. Also $\what{\delta
h}$ can not decay via renormalizable interaction terms to SM
particles.

The various decay channels of the inflaton are mainly determined
by the Lagrangian part containing two fermions -- see \Eref{mfer}.
The inflaton can decay into a pair of RH neutrinos
($\what\nu^c_{\rm I}$) through the following lagrangian terms:
\beq \label{l1}{\cal L}_{\rm I\nu_i^c} =
-\ld_{i\nu^c}{\Mpq\over\Ms}{f_0\over
J_0}\lf1-12\ck\mpq^2\rg\what{\delta
h}\what\nu^c_i\what\nu^c_i+{\rm h.c.}\,. \eeq
From \Eref{l1} we deduce that the decay of $\what{\delta h}$ into
$\what \nu^c_i$ is induced by two lagrangian terms. The first one
originates exclusively from the non-renormalizable term of
\Eref{Whi} -- as in the case of a similar model in \cref{jean}.
The second term is a higher order decay channel due to the SUGRA
lagrangian -- cf. \cref{Idecay}. The interaction in \Eref{l1}
gives rise to the following decay width
\beq \GNsn={c_{\rm I\nu^c}^2\over64\pi}\msn\sqrt{1-{4\mrh[\rm
I]^2\over\msn^2}}\>\>\>\mbox{with}\>\>\> c_{\rm I\nu^c}={\mrh[\rm
I]\over\Mpq}{f_0^{3/2}\over J_0}\lf1-12\ck\mpq^2\rg,
\label{Gpq}\eeq
where $\mrh[\rm I]$ is the Majorana mass of the RH neutrino into
which the inflaton can decay -- see \Tref{tab3}. In addition, it
was \cite{Idecay} recently recognized that within SUGRA the
inflaton can decay to the MSSM particles spontaneously -- i.e.,
even without direct superpotential couplings -- via non
renormalizable interaction terms. For a typical trilinear
superpotential term of the form $W_y=yXYZ$, we obtain the
effective interactions described by the langrangian part
\beq \label{l2} {\cal L}_{{\rm I}y} =
6y\ck{\Mpq\over\mP^2}{f_0^{3/2}\over 2J_0}\what{\delta h}\lf \what
X\what\psi_{Y}\what\psi_{Z}+\what
Y\what\psi_{X}\what\psi_{Z}+\what
Z\what\psi_{X}\what\psi_{Y}\rg+{\rm h.c.}\,, \eeq
where $y$ is a Yukawa coupling constant, $\psi_X, \psi_{Y}$ and
$\psi_{Z}$ are the chiral fermions associated with the superfields
$X, Y$ and $Z$, and whose scalar components are denoted with the
superfield symbol. For these scalars a term similar to the second
one inside $\ln$ of \Eref{Kcom} is assumed so as to obtain
canonically normalized scalars and spinors through relations
similar to the last equalities in Eqs.~(\ref{VJe}) and (\ref{y1})
respectively. Taking into account the terms of \Eref{wmssm} and
the fact that the adopted SUSY GUT predicts Yukawa unification for
the 3rd generation at $\Mpq$, we conclude that the interaction
above gives rise to the following 3-body decay width
\beq \Ghsn={14 c_{{\rm
I}y}^2\over512\pi^3}\msn^3\simeq{3y_{33}^2
\over64\pi^3}f_0^3\lf\msn\over\mP\rg^2\msn\>\>\>\mbox{where}\>\>\>
c_{{\rm I}y}=6y_{33}\ck{\Mpq\over\mP^2}{f_0^{3/2}\over J_0},
\label{Gpq1}\eeq
with $y_{33}\simeq(0.4-0.6)$ being the common Yukawa coupling
constant of the third generation computed at the $\msn$ scale and
summation is taken over color and weak hypercharge d.o.f, in
conjunction with the assumption that $\msn<2\mrh[3]$ which is
valid for both inflaton-decay scenaria considered below.

Since the decay width of the produced $\what\nu^c_{\rm I}$ is much
larger than \Gsn -- see below -- the reheating temperature,
$\Trh$, is exclusively determined by the inflaton decay and is
given by \cite{quin}
\beq \label{T1rh} \Trh=
\left(72\over5\pi^2g_{*}\right)^{1/4}\sqrt{\Gsn\mP}
\>\>\>\mbox{with}\>\>\>\Gsn=\GNsn+\Ghsn,\eeq
where $g_{*}$ counts the effective number of relativistic degrees
of freedom at temperature $\Trh$. For the MSSM spectrum,
$g_{*}\simeq228.75$. From \Fref{fig1} we remark that $\Trh$ does
not exclusively increase with $\ld$, but rather follows the
behavior of $\msn$. For $\mrh[\rm I]=10^{11}~\GeV$, we find that
$\Ghsn$ dominates over $\GNsn$ for $\ld\gtrsim0.002$ or -- via
\Eref{lan} -- $\ck\geq50$. For $0.0013\lesssim\ld\lesssim0.03$,
$\Trh$ remains almost constant since $f_0^3\simeq1$ -- see
\Eref{Gpq1}. For $\ld\gtrsim0.03$, $f_0^3\simeq1+12\ck\mpq^2$
starts to deviate from unity and so, $\Trh$ increases as shown in
\Fref{fig1}.

If $\Trh<\mrh[\rm I]$, the out-of-equilibrium condition
\cite{baryo} for the implementation of leptogenesis is
automatically satisfied. Subsequently $\what\nu^c_{\rm I}$ decay
into $H_u$ and $L_i^*$ via the tree-level couplings derived from
the second term in the RHS of Eq.~(\ref{wmssm}). Interference
between tree-level and one-loop diagrams generates a lepton-number
asymmetry $\ve_L$ \cite{baryo}, when CP conservation is violated.
The resulting lepton-number asymmetry after reheating can be
partially converted through sphaleron effects into baryon-number
asymmetry. However, the required $\Trh$ must be compatible with
constraints for the $\Gr$ abundance, $Y_{\Gr}$, at the onset of
\emph{nucleosynthesis} (BBN). In particular, the $B$ yield can be
computed as
\beq {\sf\ftn
(a)}\>\>\>Y_B=-0.35Y_L\>\>\>\mbox{with}\>\>\>{\sf\ftn
(b)}\>\>\>Y_L=2\ve_L{5\over4}
{\GNsn\over\Gsn}{\Trh\over\msn}\cdot\label{Yb}\eeq
The numerical factor in the RHS of \sEref{Yb}{a} comes from the
sphaleron effects, whereas the numerical factor ($5/4$) in the RHS
of \sEref{Yb}{b} is due to the slightly different calculation
\cite{quin} of $\Trh$ -- cf.~\cref{baryo}. In the major part of
our allowed parameter space -- see \Sref{num} -- $\Gsn\simeq\Ghsn$
and so the involved branching ratio of the produced $\what\nu_i^c$
is given by
\beq
{\GNsn\over\Gsn}\simeq{\GNsn\over\Ghsn}={\pi^2\lf1-12\ck\mpq^2\rg^2\over72\ck^2y_{33}^2\mpq^4}{\mrh[\rm
I]^2\over\msn^2}\cdot\label{Gmqq}\eeq
For $\mrh[\rm I]\simeq\lf10^{11}-10^{13}\rg\GeV$ the ratio above
takes adequately large values so that $Y_L$ is sizable. Therefore,
the presence of more than one inflaton decay channels does not
invalidate the non-thermal leptogenesis scenario. On the other hand,
the $\Gr$ yield at the onset of BBN is estimated to be \cite{kohri}:
\beq\label{Ygr} Y_{\Gr}\simeq c_{\Gr}
\Trh\>\>\>\mbox{with}\>\>\>c_{\Gr}=1.9\cdot10^{-22}/\GeV.\eeq
Let us note that non-thermal $\Gr$ production within SUGRA is
\cite{Idecay} also possible. However, we here prefer to adopt the
conservative approach based on the estimation of $Y_{\Gr}$ via
\Eref{Ygr} since the latter $\Gr$  production depends on the
mechanism of the SUSY breaking.

Both \eqs{Yb}{Ygr} calculate the correct values of the $B$ and
$\Gr$ abundances provided that no entropy production occurs for
$T<\Trh$, as we already assumed -- see \Sref{fhi2}. Regarding the
estimation of $\ve_L$, appearing in \Eref{Yb}, we single out two
cases below, according to whether the inflaton can decay into the
lightest ($\what\nu^c_1$) or to the next-to-lightest
($\what\nu^c_2$) RH neutrino. Note that the decay of the inflaton
to the heaviest RH neutrino ($\what\nu^c_3$) is disfavored by the
kinematics and the $\Gr$ constraint.

\subsubsection{Decay of the Inflaton to the Lightest RH Neutrino}\label{sc1}

In this case, we suppose that the Majorana masses of $\sni$ are
hierarchical, with $\mrh[1]\ll \mrh[2],\mrh[3]$ (but with $\mrh[1]
> \Trh$). The produced lepton-number asymmetry for a normal
hierarchical mass spectrum of light neutrinos reads \cite{baryo}
\beq\label{el} \ve_L = -\frac {3}{8\pi}\frac{\mntau
\mrh[1]}{\vev{\hu}^2} \deff\,. \eeq
Here $|\deff|\leq1$, which we treat as a free parameter,
represents the magnitude of CP violation; $\mntau$ is
the mass of heaviest light neutrino $\nu_\tau$ and we take
$\vev{\hu}=174~\GeV$ -- adopting the large $\tan\beta$ regime.

\subsubsection{Decay of the Inflaton to the Next-to-Lightest RH Neutrino}\label{sc2}

In this case, we assume $\mrh[1]\ll\mrh[2]\ll\mrh[3]$ and impose
the conditions $\Trh<\mrh[2] < \msn/2$ and $\mrh[1]>\Trh$. The
resulting lepton asymmetry is \cite{vlachos, jean}:
\beq \label{el2} \ve_L = {3\over8\pi}{\mrh[2]\over\mrh[3]}{\lf
\mD[3]^2 - \mD[2]^2\rg^2\sth^2\cth^2\sin2\delta\over\vev{H_u}^2\lf
\mD[3]^2\sth^2 + \mD[2]^2\cth^2\rg}, \eeq
where $\mD[i]$ are the Dirac masses of $\nu_i$ -- in a basis where
they are diagonal and positive -- $\cth = \cos\vartheta, \sth =
\sin \vartheta$, with $\vartheta$ and $\delta$ being the rotation
angle and phase which diagonalize the Majorana-mass matrix,
$M_{\what\nu^c}$, of $\what\nu^c_i$. The values of the various
parameters are estimated at the leptogenesis scale. Note that
since $\mrh[1]>\Trh$, $\ve_L$ calculated by \Eref{el2} is not
washed out due to $\what\nu^c_1$ inverse decays and $\Delta L=1$
scatterings -- the case with $\mrh[1]<\Trh$ is treated in
\cref{senoguz}. Also \Eref{el2} holds provided that
$\mrh[2]\ll\mrh[3]$ and the decay width of $\what\nu^c_i$ is much
smaller than $(\mrh[3]^2-\mrh[2]^2)/\mrh[2]$. Both conditions are
well satisfied here -- see Sec.~4.3.2.

Since recent results \cite{Expneutrino} -- see, also, \cref{riaz}
-- show that the mixing angle $\theta_{13}$ can be taken (at
$95\%$ c.l.) equal to zero and earlier analysis \cite{lisi} of the
CHOOZ experiment~\cite{chooz} suggests that the solar and
atmospheric neutrino oscillations decouple, we are allowed to
concentrate on the two heaviest families ignoring the first one.
This assumption enables us to connect this leptogenesis scenario
with the oscillation parameters of the $\nu_\mu-\nu_\tau$ system.
The light-neutrino mass matrix, $m_\nu$, is given by the seesaw
formula:
\beq \label{seesaw} m_\nu= -m_{\rm D}^{\tr} M_{\what\nu^c}^{-1}
m_{\rm D}, \eeq
where $m_{\rm D}$ is the Dirac mass matrix of the $\nu_i$. The
determinant and the trace invariance of $m_\nu^\dg m_\nu$ imply
two constraints on the parameters involved, i.e.
\beq \label{mDM} \mD[2]^2\mD[3]^2=
\mntau\mnmu\mrh[2]\mrh[3]\>\>\>\mbox{and}\>\>\>A_{\rm
D}\sth^2+B_{\rm D}\sth+C_{\rm D}=0\eeq with the coefficient of the
latter equation being
\beqs\baq \nonumber A_{\rm D}&=&(\mD[2]^2 -
\mD[3]^2)\Big((\mD[2]^2- \mD[3]^2)\mrh[2]^2 + (\mD[2]^2 - \mD[3]^2)\mrh[3]^2\\
&+&2(\mD[2]^2 + \mD[3]^2)\mrh[2]\mrh[3]\cos2\delta\Big)
\\ \nonumber B_{\rm D}&=&2(\mD[2]^2 - \mD[3]^2)\Big(\mD[3]^2\mrh[2]^2 -\mD[2]^2\mrh[3]^2 \\ &-&
(\mD[2]^2 +\mD[3]^2) \mrh[2]\mrh[3]\cos2\delta\Big)
\\ C_{\rm D}&=&\lf\mD[2]^4\mrh[3]^2 + \mD[3]^4\mrh[2]^2\rg - \lf\mntau^2 +
\mnmu^2\rg\mrh[2]^2\mrh[3]^2. \eaq\eeqs
Assuming that the Dirac mixing angle (i.e., the mixing angle in
the absence of RH neutrino Majorana masses) is negligible, we can
identify~\cite{vlachos} the rotation angle which diagonalizes
$m_{\nu}$ with the physical mixing angle in the $\nu_\mu-\nu_\tau$
leptonic sector, $\theta_{23}$, which is constrained by the
present neutrino data -- see below.

\subsection{Imposed Constraints}\label{cont1}

The parameters of our model can be further restricted if, in
addition to the inflationary requirements mentioned in \Sref{fhi2}
and the restriction $\ld\geq0.001$ which allows us to ignore
effects of instant preheating -- see Appendix B -- we impose extra
constraints arising from the post-inflationary scenario. These are
the following:

\paragraph{4.2.1} To ensure that the inflaton decay to RH
neutrinos is kinematically allowed we have to impose the
constraint:
\beq\label{kin} \msn\geq2\mrh[\rm
I]\>\>\>\Rightarrow\>\>\>\mrh[\rm
I]\lesssim\ld\mP/4\sqrt{3}\ck\simeq1.5\cdot10^{13}~\GeV\>\>\>\mbox{for}\>\>\>\ld\gtrsim1.3\cdot10^{-3},\eeq
where we make use of \Eref{mqa}. More specifically, we require
$\msn\geq2\mrh[1]$ [$\msn\geq2\mrh[2]$] for the scenario described
in \Sref{sc1} [\Sref{sc2}].

\paragraph{4.2.2} In agreement with our assumption about hierarchical light
neutrino masses for both inflaton-decay scenaria, the solar and
atmospheric neutrino mass squared differences $\Delta m^2_\odot$
and $\Delta m^2_\oplus$ can be identified with the squared masses
of $\nu_\mu$ and $\nu_\tau$, $\mnmu^2$ and $\mntau^2$, respectively.
Taking the central values of the former quantities
\cite{Expneutrino}, we set:
\beq\label{mmt} {\sf\ftn (a)}\>\>\>\mnmu=\sqrt{\Delta
m^2_\odot}=0.0087\>{\rm eV}\>\>\>\mbox{and}\>\>\>{\sf\ftn
(b)}\>\>\>\mntau=\sqrt{\Delta m^2_\oplus}=0.05\>{\rm eV}.\eeq
We multiply the values above by $1.12$ in order to roughly
approximate renormalization group effects for the evolution of
these masses from the electroweak up to the leptogenesis scale --
see Fig.~4 of \cref{runing}. The resulting $\mntau(\Trh)$ is low
enough to ensure that the lepton asymmetry is not erased by lepton
number violating $2\to2$ scatterings \cite{erasure} at all
temperatures between $\Trh$ and $100~\GeV$. Also
$\sin^2\theta_{23}$ is to be consistent with the following $95\%$
c.l. allowed range \cite{Expneutrino}:
\beq 0.41\lesssim\sin^2\theta_{23}\lesssim0.61. \label{s8}\eeq

\paragraph{4.2.3} Due to the presence of $SU(4)_{\rm C}$ symmetry in $\Ggut$,
$\mD[3]$ coincides with the value of the top quark mass, $m_t$, at
the leptogenesis scale -- see also \Eref{wmssm} -- i.e.
\beq\label{mtop}\mD[3](\Trh)=m_t(\Trh).\eeq
Working in the context of the MSSM with $\tan\beta=50$, and
solving the relevant renormalization group equations, we find
$m_t(\Trh)\simeq\lf120-126\rg~\GeV$ for the values of $\Trh$ encountered
in our set-up.

\paragraph{4.2.4} The implementation of BAU via non-thermal
leptogenesis dictates \cite{wmap} at 95\% c.l.
\beq Y_B=\lf8.74\pm0.42\rg\cdot10^{-11}\>\Rightarrow\>8.32\leq
Y_B/10^{-11}\leq9.16.\label{BAUwmap}\eeq

\paragraph{4.2.5} In order to avoid spoiling the success of the
BBN, an upper bound on $Y_{\Gr}$ is to be imposed depending on the
$\Gr$ mass, $m_{\Gr}$, and the dominant $\Gr$ decay mode. For the
conservative case where $\Gr$ decays with a tiny hadronic
branching ratio, we have \cite{kohri}
\beq  \label{Ygw} Y_{\Gr}\lesssim\left\{\bem
10^{-15}\hfill \cr
10^{-14}\hfill \cr
10^{-13}\hfill \cr
10^{-12}\hfill \cr\eem
\right.\>\>\>\mbox{for}\>\>\>m_{\Gr}\simeq\left\{\bem
0.45~{\rm TeV}\hfill \cr
0.69~{\rm TeV}\hfill \cr
10.6~{\rm TeV}\hfill \cr
13.5~{\rm TeV.}\hfill \cr\eem
\right.\eeq
The bound above can be somehow relaxed in the case of a stable
$\Gr$. As we see below, this bound is achievable  within our model
model only for $m_{\Gr}>10~\TeV$. As $m_{\Gr}$ gets larger than
this bound leads to the necessity of a rather fine tuned SUSY
breaking mechanism such as split SUSY \cite{split}, or anomaly
mediated SUSY breaking \cite{randal} where the superpartners of SM
particles have masses lower than $m_{\Gr}$. Using \Eref{Ygr} the
bounds on $Y_{\Gr}$ can be translated into bounds on $\Trh$.
Specifically we take $\Trh\simeq\lf0.53-5.3\rg\cdot10^8~\GeV$
[$\Trh\simeq\lf0.53-5.3\rg\cdot10^9~\GeV$] for
$Y_{\Gr}\simeq\lf0.1-1\rg\cdot10^{-13}$
[$Y_{\Gr}\simeq\lf0.1-1\rg\cdot10^{-12}$].

\subsection{Numerical Results}\label{num}

Considering the constraints above in conjunction with those quoted
in \Sref{fhi2}, we can delineate the overall allowed parameter
space of our model. Recall that we set $\lH=\lHb=0.5,\, \kx=1$ and
$g=0.7$ with $\Mpq$ given by \Eref{Mpqf}. The upper [lower] bound
of the used $\ld$'s comes from \Eref{res1} [the conclusions of
Appendix B]. Also we fix $y_{33}=0.5$ throughout our
investigation. As can be deduced by \eqs{Yb}{Gmqq}, for lower
[larger] $y_{33}$'s satisfying \Eref{BAUwmap} requires slightly
[lower] larger $\mrh[\rm I]$'s. As mentioned in \Sref{lept}, we
adopt two leptogenesis scenaria depending on whether $\what\dth$
can decay to $\what\nu^c_1$ or $\what\nu^c_2$. As we see below, in
both cases, two disconnected allowed domains arise according to
which of the two contributions in \Eref{l1} dominates. The
critical point $(\ld_{\rm c}, c_{\cal R\rm c})$ is extracted from:
\beq 1-12c_{\cal R\rm c}\mpq^2=0\Rightarrow c_{\cal R\rm
c}=1/12\mpq^2\>\>\mbox{or}\>\>\ld_{\rm
c}\simeq10^{-4}/25.2\mpq^2\simeq0.06\eeq
where we make use of \Eref{lan} in the last step. From
\eqss{T1rh}{Yb}{Gmqq} one can deduce that for $\ld<\ld_c$, $\Trh$
remains almost constant; $\GNsn/\Gsn$ decreases as $\ck$ increases
and so the $\mrh[\rm I]$'s, which satisfy \Eref{BAUwmap},
increase. On the contrary, for $\ld>\ld_c$, $\GNsn/\Gsn$ is
independent of $\ck$ but $\Trh$ increases with $\ck$ and so
fulfilling \Eref{BAUwmap} $\mrh[\rm I]$'s decrease.

In the following, we present the results of our investigation in
each case, separately.

\paragraph{4.3.1. Decay of the inflaton into {\normalsize
$\widehat{\nu}^c_1$}.}

In this case our cosmological setting depends on the following
parameters:
$$\ld,\>\ck,\>\mrh[1]\>\>\>\mbox{and}\>\>\>\deff.$$
Given our ignorance about $\deff$ in \Eref{el}, we take $\deff=1$,
allowing us to obtain via \Eref{el} the maximal \cite{Ibara}
possible $|\eL|$. This choice in conjuction with the imposition of
the lower bound on $\Yb$ in \Eref{BAUwmap} provides the most
conservative restriction on our parameters.

\begin{figure}[!t]\vspace*{-.46in}\begin{tabular}[!h]{cc}\begin{minipage}[t]{7in}
\hspace{0.5in}
\epsfig{file=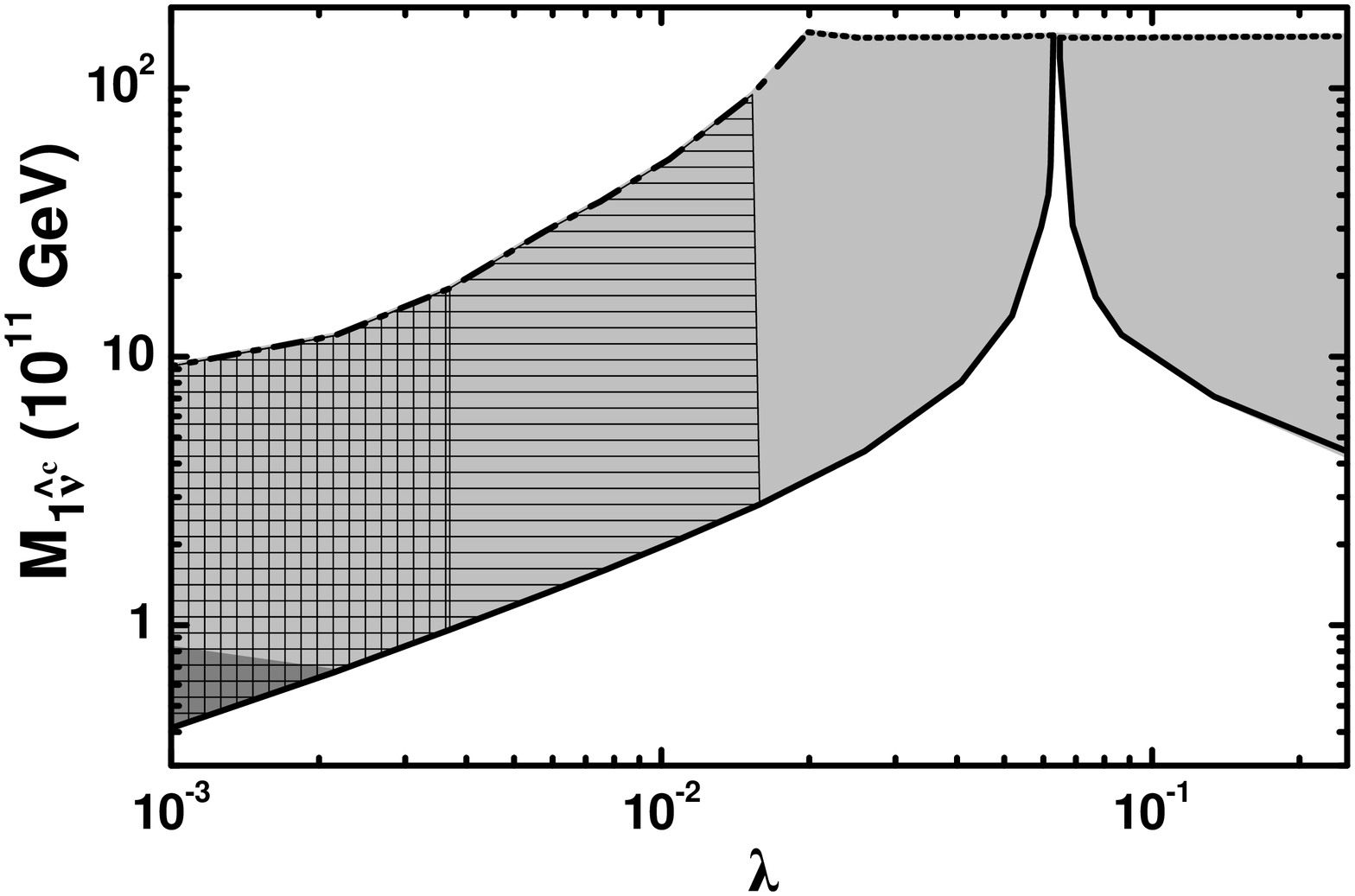,height=3.65in,angle=-90}\end{minipage}
&\begin{minipage}[h]{3in}
\hspace{-3.in}{\vspace*{-2.6in}\includegraphics[height=4.3cm]
{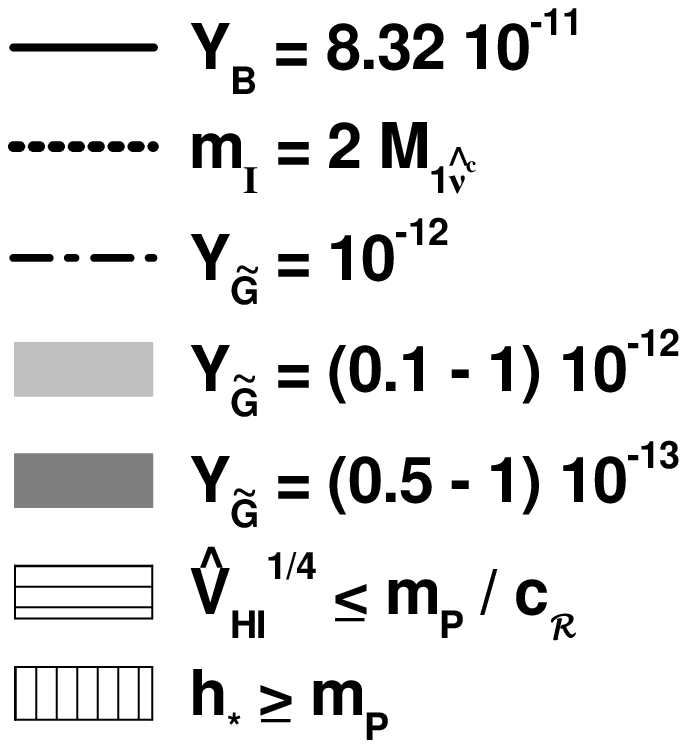}}\end{minipage}
\end{tabular} \hfill \vchcaption[]{\sl \small Allowed (shaded) regions
in the $\ld-\mrh[1]$ plane, for $\lH=\lHb=0.5, \kx=1$ and
$y_{33}=0.5$ when the inflaton can decay into $\what\nu^c_1$'s.
The conventions adopted for the various lines and shaded or
hatched regions are also shown.}\label{fig2}
\end{figure}

As we explain in \Sref{num1}, $\mrh[1]$ does not crucially
influence the inflationary observables. On the contrary, $\mrh[1]$
plays a key-role in simultaneously satisfying
\eqss{kin}{BAUwmap}{Ygw} -- see \eqs{Yb}{Ygr}. For this reason we
display in \Fref{fig2} the allowed regions by all the imposed
constraints in the $\ld-\mrh[1]$ plane. In the horizontally lined
regions \sEref{Vl}{a} holds, whereas in the vertically hatched
region we get $\sg_*\geq\mP$. The restrictions on the parameters
arising from the post-inflationary era are depicted by solid,
short dotted and dot-dashed lines. Namely, the solid [dot-dashed]
lines correspond to the lower [most conservative upper] bound on
$Y_B$ [$\Yg$] in \Eref{BAUwmap} [\Eref{Ygw}]. Since we use
$|\deff|=1$, it is clear from \eqs{Yb}{el} that values of
$\mrh[1]$ above the solid line are compatible with the current
data in \Eref{BAUwmap} for conveniently adjusing $|\deff|\leq1$.
However the values of $\mrh[1]$ can be restricted by the bounds of
\Eref{Ygw}. Specifically we obtain
$Y_{\Gr}\simeq\lf0.1-1\rg\cdot10^{-12}$
[$Y_{\Gr}\simeq\lf0.5-1\rg\cdot10^{-13}$]  in the [gray] dark gray
area. On the other hand, the kinematical condition depicted by a
short dotted line -- see \Eref{kin} -- puts the upper bound on
$\mrh[1]$ in a the upper right corners of the allowed region. As
anticipated above, this region has two disconnected branches. In
the left [right] branch, the first [second] term in \Eref{l1}
dominates. All in all we obtain
\beqs\bea\label{res2a}
0.39\lesssim\mrh[1]/10^{11}~\GeV\lesssim154\>\>\>
\mbox{for}\>\>\>0.001\lesssim\ld\lesssim0.062,\\
154\gtrsim\mrh[1]/10^{11}~\GeV\gtrsim4.32\>\>\>
\mbox{for}\>\>\>0.062\lesssim\ld\lesssim0.25.
\label{res2b}\eea\eeqs
The overall minimal [maximal] $\mrh[1]$ can be found in the left
lower [upper right] corner of the allowed region. Obviously, the
maximal allowed $\mrh[1]\simeq\msn/2$ is obtained for
$0.02\lesssim\ld\lesssim0.062$ and $0.062\lesssim\ld\lesssim0.25$.
This region gives also a lower bound on $|\deff|$,
$|\deff|\gtrsim1.6\cdot10^{-4}$.


Trying to compare, finally, the resulting allowed parameter space
here in the $\ld-\mrh[1]$ plane with the one allowed within the
models of HI \cite{susyhybrid} -- see Fig. 2 of \cref{jeanl},
where the coupling constant $\kappa$ corresponds to $\ld$ -- we
remark that in our case {\sf (i)} higher $\ld$'s and $\mrh[1]$'s
are allowed and {\sf (ii)} there is an additional minor slice of
allowed parameters exclusively due to the SUGRA induced decay
channel found in \Eref{l1}.

\paragraph{4.3.2. Decay of the inflaton into {\normalsize $\widehat{\nu}^c_2$}. }

\begin{figure}[!t]\vspace*{-.3in}
\begin{center}
\epsfig{file=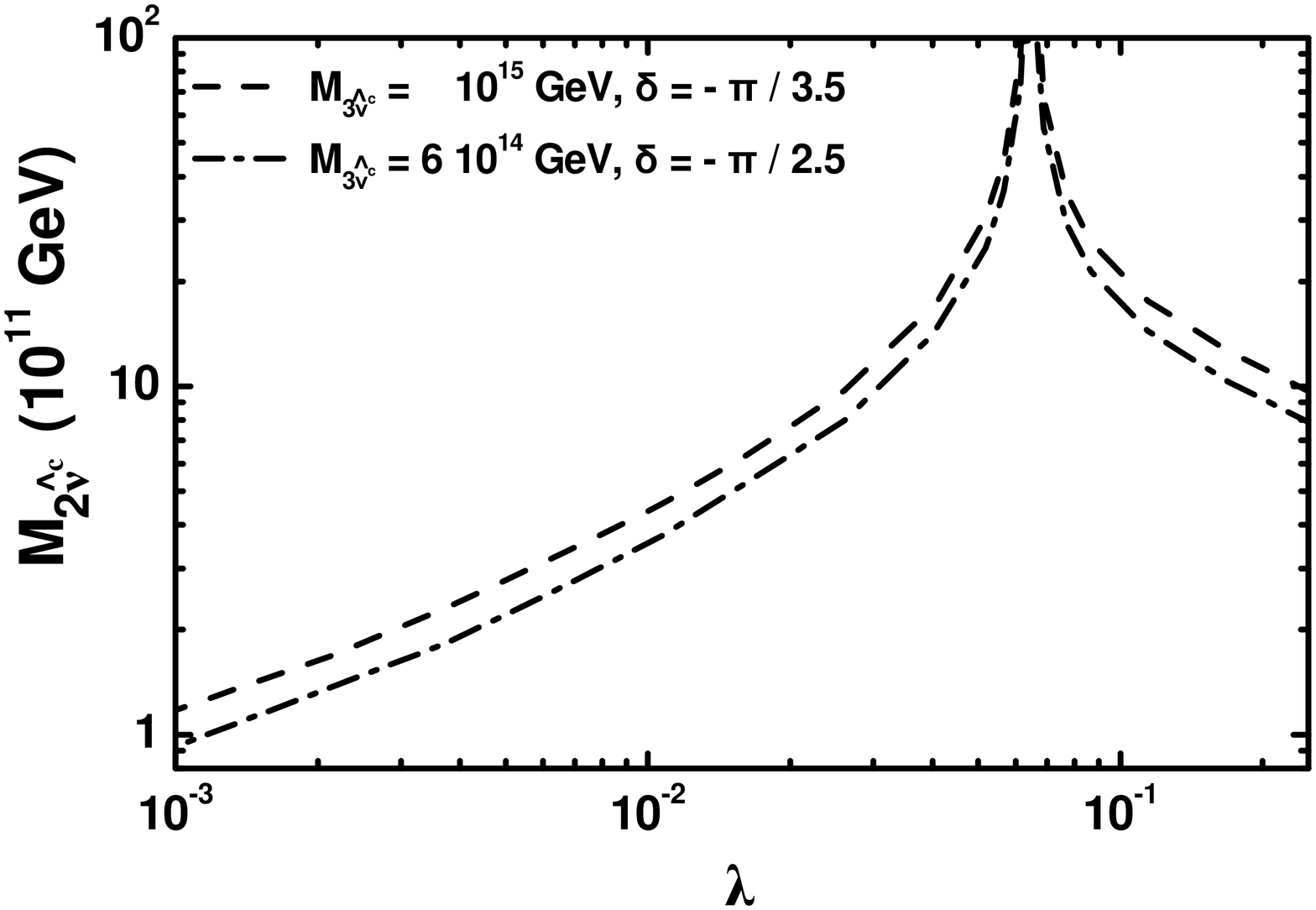,height=3.65in,angle=-90}
\end{center}
\renewcommand{\arraystretch}{1.1}
\begin{center} \begin{tabular}{|c|c||c|c|c|c|c|} \hline
\multicolumn{2}{|c||}{ \small\sc Input Parameters}
&\multicolumn{5}{|c|}{ \small\sc Output Parameters}\\ \hline
\hline
$\mrh[3]$&$-\delta$&$\ld~(10^{-2})$&$\mrh[2]$
&$\sin^2\theta_{23}$&$\vartheta~(10^{-2})$&$\mD[2]~(\GeV)$\\
$(\GeV)$&&&$ (10^{11}~\GeV) $&$(10^{-2})$&&\\
\hline
$10^{15}$&$\pi/3.5$&$0.1-6.2$ &$1.5-132$&$45.7-46.5$&$1.2-13$&$1.9-21.1$\\
&&$6.6-25$&$9.5-142$&$44-46$&$3.6-1.4$&$5.7-21.9$\\
$6\cdot 10^{14}$&$\pi/2.5$&$0.1-6.2$&$0.9-97$&$44.8-45.6$&$2-17$&$1.5-14$\\
&&$6.6-25$&$7.7-94$&$45.3-47$&$4.7-12$&$4-10$\\
\hline
\end{tabular}\end{center} \hfill \vchcaption[]{\sl \small Allowed values of $\mrh[2]$
versus $\ld$, for the input quantities listed in the table above,
$\lH=\lHb=0.5, \kx=1$ and $y_{33}=0.5$ when the inflaton can decay
into $\widehat{\nu}^c_2$'s. The conventions adopted for the types
and the color of the various lines are also shown.}\label{fig3}
\end{figure}

In this case, our cosmological setting depends on the following
input parameters:
$$\ld,\>\ck,\>\mrh[2],\>\mrh[3]\>\>\>\mbox{and}\>\>\>\delta.$$
In our numerical program, for every $\ld$ and $\ck$ consistent
with the inflationary requirements of \Sref{fhi2}, we can resolve
\Eref{mDM} w.r.t $\mD[2]$ and $\vartheta$, if we use $\mrh[2]$ and
$\delta$ as input parameters -- recall that $\mD[3]$ is determined
by \Eref{mtop}. Diagonalizing $m_\nu$ and employing \eqs{Yb}{el2}
to estimate $Y_B$, we can restrict the parameters through
\eqs{s8}{BAUwmap}. In order to compare the value of
$\sin^2\th_{23}$ extracted at the leptogenesis scale, $\Trh$, with
the low energy experimental result of \Eref{s8}, we solve the
relevant renormalization group equations following \cref{runing}.
We remark that $\sin^2\th_{23}$ increases by almost $8\%$ due to
these renormalization effects.

In \Fref{fig3} we display the allowed values of $\mrh[2]$ versus
$\ld$ for various $\mrh[3]$ and $\delta$'s indicated on the left
upper corner of the graph. The obtained allowed ranges of several
other quantities involved are arranged in the Table below the
graph. Along the displayed curves the central value of $Y_B$ in
\Eref{BAUwmap} is achieved and we obtain
$\Trh\simeq\lf0.7-2.9\rg\cdot10^{9}~\GeV$ which is translated as
$Y_{\Gr}\simeq\lf1.3-5.5\rg\cdot10^{-13}$ via \Eref{Ygr}.  From
our results we observe that increasing $\mrh[3]$ entails an
increase of $\mrh[2]$ too whereas $-\delta$ approaches $\pi/4$ --
see \Eref{el2} -- and increases as $\mrh[2]$ drops in order
\Eref{s8} to be met. The resulting $\mD[2]$ turns out to be in the
range $(1.5-21.9)~\GeV$ which is larger than the mass of the charm
quark at $\Trh$, $m_c\simeq0.427~\GeV$. Therefore within our
scheme the $SU(4)_{\rm C}$ symmetry does not hold in the up sector
of the second family.

As for the case of Sec.~4.3.1, we obtain two separate branches of
allowed parameters, $h_*\geq\mP$ for $\ld\leq0.0037$, whereas
\sEref{Vl}{a} is satisfied for $\ld\leq0.016$ -- \sEref{Vl}{b} is
valid for the used $\ld$'s, see \Sref{num1}. Finally, it is
remarkable that the assumptions on $\mrh[1]$,
$\Trh<\mrh[1]\ll\mrh[2]$ can be fulfilled in a wide range, i.e.,
$7\cdot10^{8}<\mrh[1]/\GeV<10^{11}$.

\section{Conclusions}\label{con}

In this paper we attempted to embed within a realistic GUT, based
on the PS gauge group, one of the recently formulated
\cite{linde2} SUSY models of chaotic inflation with non-minimal
coupling to gravity. We showed that the model not only supports
non-MHI driven by the radial component of the Higgs field, but it
also leads to the spontaneous breaking of the PS gauge group to
the SM one with the GUT breaking v.e.v identified with the SUSY
GUT scale and without overproduction of monopoles. Moreover,
within our model, we can resolve the strong CP and the $\mu$
problems of the MSSM via a Peccei-Quinn symmetry breaking
transition. Inflation is followed by a reheating phase, during
which the inflaton can decay into the lightest or the
next-to-lightest RH neutrino allowing, thereby for non-thermal
leptogenesis to occur via the subsequent decay of the RH
neutrinos. Although other decay channels to the SM particles via
non-renormalizable interactions are also activated, we showed
that, in both cases, the production of the required by the
observations BAU can be reconciled with the observational
constraints on the inflationary observables and the $\Gr$
abundance, provided that the (unstable) $\Gr$ masses are greater
than $10~{\rm TeV}$. In the first inflaton-decay scenario, we
restrict the lightest RH neutrino mass to values of the order
$\lf10^{11}-10^{13}\rg~\GeV$ whereas in the second scenario, extra
restrictions from the light neutrino data and the $SU(4)_{\rm C}$
factor of the adopted GUT gauge group can be also met for masses
of the heaviest [next-to-lightest] RH neutrino of the order of
$10^{15}~\GeV$ $\left[\lf10^{11}-10^{13}\rg~\GeV\right]$.

Finally, we would like to point out that, although we have
restricted our discussion on the PS gauge group, non-MHI analyzed
in our paper has a much wider applicability. It can be realized
within other GUTs, which may (as in the case of $\Ggut$) or may
not lead to the formation of cosmic defects. If we adopt another
GUT gauge group, the inflationary predictions and the
post-inflationary evolution are expected to be quite similar to
the ones discussed here with possibly different analysis of the
stability of the inflationary trajectory, since different Higgs
superfield representations may be involved in implementing gauge
symmetry breaking to $G_{\rm SM}$ -- see, e.g., \cref{su5h} which
appeared when this work was under completion.

\begin{acknowledgement} \paragraph{}

We would like to cordially thank G. Lazarides, Q. Shafi and J.
Vergados for helpful discussions and Y. Watanabe for an enlightening
correspondence.

\end{acknowledgement}
\appendix
\setcounter{equation}{0}
\renewcommand{\theequation}{A.\arabic{equation}}
\renewcommand{\thesubsubsection}{A.\arabic{subsubsection}}
\section*{Appendix A: Derivation of the Mass Spectrum During
non-MHI} \label{sugra}

\rhead[\fancyplain{}{ \bf \thepage}]{\fancyplain{}{\sl Non-MHI \&
non-Thermal Leptogenesis in a SUSY PS Model}}
\lhead[\fancyplain{}{\sl
\hspace*{-.3cm}\leftmark}]{\fancyplain{}{\bf \thepage}} \cfoot{}

In this Appendix, we describe the derivation of the mass spectrum of
the model when the radial component, $h$, of the fields $\nu^c_H$,
$\bar{\nu}^c_H$ slowly rolls down $\Vhi$, breaking $\Ggut$ down to
$G_{\rm SM}$. We explain the results summarized in
\Tref{tab2}, working exclusively in the EF. We demonstrate below
the origin of the masses of the scalars (\Sref{mscalars}),
gauge bosons (\Sref{mgbosons}) and fermions (\Sref{mfermions}).

\subsubsection{Masses for the Scalars \label{mscalars}}

Expanding $\Vhi$ in \Eref{Sni} to quadratic order in the fluctuations around
the trajectory in
\Eref{inftr}, for given $h$, we obtain
\bea \Vhi&=&\Vhio+{1\over2}m_{\widehat
S}^2\widehat{S}^2+{1\over2}m_{\widehat
\th_\nu}^2\widehat{\th}_\nu^2+{1\over2} \lin{\widehat
\th}{\widehat{\bar \th}} M_\th^2\stl{\widehat \th}{\widehat{\bar
\th}}\nonumber \\ &+&{1\over2}\sum_{x}\lf\lin{\widehat
x_1}{\widehat{\bar x}_1} M_{x1}^2\stl{\widehat x_1}{\widehat{\bar
x}_1}+\lin{\widehat x_2}{\widehat{\bar x}_2} M_{x2}^2\stl{\widehat
x_2}{\widehat{\bar x}_2}\rg, \label{Vexp}\eea
where $x=u,e,d$ and $g$ and the decomposition of the scalar fields
into real and imaginary parts is shown in \Eref{cannor}. The
various mass-squared matrices involved in \Eref{Vexp} are found to
be
\beqs \bea \label{M1} &M^2_\th=
\mtt{1}{1}{1}{1}{m_{\what\th+}^2/2},\>\>\>
\>\>\>M^2_{g1}=M^2_{g2}=\mtt{m^2_{\widehat{g}}}{0}{0}{m^2_{\widehat{\bar
g}}},\>\>\>M^2_{y1}=\mtt{m^2_{\widehat{y}1}}{m^2_{\widehat
y2}}{m^2_{\widehat y2}}{m^2_{\widehat
y1}},&\>\>\>\> \\
& M^2_{y2}=\mtt{m^2_{\widehat{y}1}}{-m^2_{\widehat
y2}}{-m^2_{\widehat y2}}{m^2_{\widehat
y1}},\>\>\>M^2_{d1}=\mtt{m^2_{\widehat{d}1}}{m^2_{\widehat
d2}}{m^2_{\widehat d2}}{m^2_{\widehat d3}}\>\>\>\mbox{and}
\>\>\>M^2_{d2}=\mtt{m^2_{\widehat{d}1}}{-m^2_{\widehat
d2}}{-m^2_{\widehat d2}}{m^2_{\widehat d3}}& \label{M2}\eea \eeqs
where $y=u$ and $e$ and $m^2_{\what\th+}, m^2_{\widehat g}$,
$m^2_{\widehat{\bar g}}$ and also $m^2_{\widehat
S}$ are presented in \Tref{tab2}. The
elements of the remaining matrices above are found to be
\beqs\bea \label{my} &m^2_{\widehat y1}={\mP^2\xsg^2(3g^2f
+\ld^2\xsg^2)/24f^2},\>\>\>m^2_{\widehat
y2}=-{\mP^2\xsg^2(\ld^2(\xsg^2-6) + 3g^2f)/24f^2}, \\
& m^2_{\widehat d1}={\mP^2\xsg^2(\ld^2\xsg^2 +
24\lH^2f)/24f^2},\>\>\>m^2_{\widehat
d2}=-{\ld^2\mP^2\xsg^2(\xsg^2-6)/24f^2}, \label{md} \eea\eeqs
whereas $m^2_{\what d3}$ is the same as $m^2_{\what d1}$ but with
$\lH$ replaced by $\lHb$. To simplify our formulae below,
we take $\lH\simeq\lHb$. The various masses squared in
\eqs{M1}{M2} originate mainly from $\what V_{\rm HF}$ in \Eref{Vsugra}.
Additional contributions from $V_{\rm HD}/f^2$ in \Eref{Vd1} arise for
$m^2_{\what\th+}$ and $m^2_{\widehat y2}$. The orthogonal matrix
$U_K$, which diagonalizes $M_K$ in \Eref{diagMk},
diagonalizes $M^2_\th$ too. We find the following eigenstates
\beq\widehat\th_{\pm}={1\over\sqrt{2}}\lf\widehat\thb\pm\widehat\th\rg
\eeq
with eigenvalues $m_{\what \th+}^2$ and $m_{\what \th-}^2=0$
respectively. The second eigenstate corresponds to the Goldstone
mode absorbed by $A^\perp$ -- see \Sref{mgbosons} -- via the
Higgs mechanism. Upon diagonalization of $M^2_{x1}$ and $M^2_{x2}$
with $x=u,e$ and $d$, we obtain the eigenvalues
\beq m^2_{\what x\pm}=m^2_{\what x1}\pm m^2_{\what x2}\eeq
which correspond to the following eigenstates respectively:
\beq\widehat x_{1\pm}={1\over\sqrt{2}}\lf\widehat{\bar
x}_1\pm\widehat x_1\rg\>\>\>\mbox{and}\>\>\>\widehat
x_{2\mp}={1\over\sqrt{2}}\lf\widehat{\bar x}_2\mp\widehat
x_2\rg\>\>\>\mbox{with}\>\>\>x=u^\ca,e\>\>\mbox{and}\>\>d.\eeq
Note that $m^2_{\what x+}=m^2_{\what x0}=\lambda^2m^2_Px^2_h/4f^2\ll m^2_{\what x-}$ for
$x=u$ (3 colors) and $x=e$ are the masses
squared of the Goldstone bosons $\what x_{1+}$ and $\what x_{2-}$
which are absorbed by $A_{\rm C}^\pm$ (3 colors) and $A_{\rm R}^\pm$
-- see \Sref{mgbosons} -- via the Higgs mechanism. The
remaining $m^2_{\what x\pm}$'s are listed in \Tref{tab2}.

\subsubsection{Masses for the Gauge Bosons \label{mgbosons}} Some of the gauge bosons
$A_{\rm C}^a$ and $A_{\rm R}^m$ acquire masses from the
lagrangian terms -- cf. \cref{oliveira}:
\bea \nonumber & K_{\al\bbet} \lf\lf D_\mu \bar H^{c*}\rg^\al\lf
D^\mu \bar H^{c \tr}\rg^\bbet + \lf D_\mu H^{c\dg}\rg^\al \lf D^\mu
H^c\rg^\bbet \rg = \\ &
{1\over2}m_{\pm}^2\bigg(\lf\sqrt{\frac{3}{2}} A_{\rm
C\mu}^{15}-A_{\rm R\mu}^3\rg\lf\sqrt{\frac{3}{2}} A_{\rm
C}^{15\mu}-A_{\rm R}^{3\mu}\rg+2A_{\rm R\mu}^{+}A_{\rm
R}^{\mu-}+2\sum_{\ca=1}^3A_{\rm C\mu}^{\ca+}A_{\rm
C}^{\mu\ca-}\bigg), \label{cov} \eea
where $m_{\pm}$ are given in \Tref{tab2}. The action of $D_\mu$ on
$\Hcc$ and $\bHc$ is as follows
\beqs\bea D_\mu \bar H^{c \tr}&=\partial_\mu\bar H^{c \tr}+ig\lf
\sum_{a=1}^{15}T_{\rm C}^a A_{\rm C}^a\bar H^{c \tr}+
\sum_{m=1}^{3}T_{\rm R}^m A_{\rm R}^m\bar H^{c \tr}\rg\\
D_\mu H^{c}&=\partial_\mu H^{c}-ig\lf \sum_{a=1}^{15} T_{\rm
C}^{a*}A_{\rm C}^a H^{c}+ \sum_{m=1}^{3} T_{\rm R}^{m*}A_{\rm R}^m
H^{c}\rg,\eea\eeqs
and we have defined the following normalized gauge fields:
\beq \label{Apm} A_{\rm C}^{\ca\pm}={1\over\sqrt{2}}\lf A_{\rm
C}^{7+2\ca}\pm i A_{\rm
C}^{8+2\ca}\rg\>\>\>\mbox{for}\>\>\>\ca=1,2,3\>\>\>\mbox{and}\>\>\>A_{\rm
R}^{\pm}={1\over\sqrt{2}}\lf A_{\rm R}^{1}\pm i A_{\rm
R}^{2}\rg.\eeq
The first term of the RHS of \Eref{cov} can be written as
\beq {1\over2}m^2_{\pm}\lin{A_{\rm C}^{15\mu}}{A_{\rm R}^{3\mu} }
M_{\rm CR}\stl{A_{\rm C}^{15\mu}}{A_{\rm R}^{3\mu}
}={1\over2}m^2_{\perp}A^{\perp\mu}A^{\perp}_\mu~, \label{cov1}\eeq
where $m^2_{\perp}=5m^2_{\pm}/2$ -- see \Tref{tab2} -- since
\beq M_{\rm
CR}=\mtt{3/2}{-\sqrt{3/2}}{-\sqrt{3/2}}{1}\>\>\>\mbox{and}
\>\>\>U_{\rm CR}M_{\rm CR}U_{\rm CR}^\tr=\diag\lf5/2,0\rg\eeq with
\beq U_{\rm
CR}=\mtt{-\sqrt{3/5}}{\sqrt{2/5}}{\sqrt{2/5}}{\sqrt{3/5}}\>\>\>
\mbox{and}\>\>\>\stl{A^{\perp}}{A^{||}}=U_{\rm CR}\stl{A_{\rm
C}^{15}}{A_{\rm R}^{3}}\cdot \eeq
Therefore, from \eqs{cov}{cov1}, we can infer that 9 gauge bosons
($A^{\perp}$, $A_{\rm C}^{\ca\pm}$ and $A_{\rm R}^{\pm}$) become
massive, absorbing the massless modes $\what\th_-, \what
u^\ca_{1+}$, $\what u^\ca_{2-}$, $\what e_{1+}$ and $\what
e_{2-}$, whereas $A^{||}$, which remains massless, can be
interpreted as the $B$ boson associated with the $U(1)_Y$ factor of $G_{\rm SM}$.

\subsubsection{Masses for the Fermions \label{mfermions}}

The lagrangian kinetic terms of the chiral fermions $\psi_x$ with
$x=\nu, e, u, d,  g, \bar \nu, \bar e, \bar u, \bar d, \bar g$,
which are associated with the superfields $\snH, \eH, \uH, \dH,
g^c, \snHb, \eHb, \uHb, \dHb, \bar g^c$ respectively, are
\bea \nonumber K_{\al\bbet}\ovl \psi^\al \bar\sigma^\mu D_\mu
\psi^\bbet&=&  {1\over f^2}\lin{\ovl\psi_\nu}{\ovl\psi_{\bar\nu}}
M_K \stl{\psi_\nu}{\psi_{\bar\nu}}+\sum_y{1\over f}\ovl
\psi_y \bar\sigma^\mu D_\mu \psi_y \\
&=&\ovl{\what\psi}_{\nu+} \bar\sigma^\mu
D_\mu\what\psi_{\nu+}+\ovl{\what\psi}_{\nu-} \bar\sigma^\mu
D_\mu\what\psi_{\nu-}+\sum_y \ovl{\what\psi}_y  \bar\sigma^\mu
D_\mu\what\psi_y, \label{yn}\eea
where we have used \Eref{diagMk}, and the normalized spinors
are defined as follows
\beq\label{y1} \what\psi_{\nu+}={\sqrt{\bar f}\over
f}\psi_{\nu+},\>\>\>\what\psi_{\nu-}={\psi_{\nu-}\over\sqrt{f}}\>\>\>\mbox{with}\>\>\>
\psi_{\nu\pm}={1\over\sqrt{2}}\lf\psi_{\bar\nu}\pm\psi_\nu\rg\>\>\>
\mbox{and}\>\>\>\what\psi_y={\psi_y\over\sqrt{f}}\eeq
with $y=e, u, d,  g, \bar e, \bar u, \bar d, $ and $\bar g$. In
\Eref{yn} the contraction between two Weyl spinors is suppressed;
the Pauli matrices $\bar\sigma^\mu$ and the action of $D_\mu$ on
$\psi^\al$ are specified in \cref{sugraL, linde1}.

Having defined the normalized spinors, we can proceed with the
derivation of the fermionic mass spectrum of our model. The masses of
the chiral fermions can be found applying the formula
\cite{sugraL, linde1}:
\beq \label{mfer} m_{\al\bt}=e^{K/2\mP^2}\lf W_{\rm
HPS\al\bt}+{1\over\mP^2}\lf K_{\al\bt}W_{\rm
HPS}+K_{\al}F_\bt+K_{\bt}F_\al\rg-\Gamma^\gamma_{\al\bt}F_\gamma\rg\eeq
with $\Gamma^\gamma_{\al\bt}=K^{\gamma\bar\gamma}\partial_\al
K_{\bt\bar\gamma}$, $W_{\rm HPS\al\bt}=W_{{\rm
HPS},\phi^\al\phi^\bt}$, $K_{\al}=K_{,\phi^\al}$ and $F_\al$ as
defined below \Eref{Vsugra}. Upon diagonalization of the relevant
mass matrix, we obtain the eigenvalues $m_{\what\psi_{S\nu}},
m_{\what\psi_{\bar gd}}$ and $m_{\what\psi_{g\bar d}}$, listed in
\Tref{tab2}, corresponding to the following eigenstates
\beq\widehat \psi_{S\nu\pm}={1\over\sqrt{2}}\lf\widehat
\psi_S\pm\widehat \psi_{\nu+}\rg,\>\>\> \what\psi^\ca_{\bar g
d\pm}={1\over\sqrt{2}}\lf\widehat \psi^\ca_{\bar g}\pm\widehat
\psi^\ca_{d}\rg\>\>\>\mbox{and}\>\>\>\what\psi^\ca_{g\bar
d\pm}={1\over\sqrt{2}}\lf\widehat \psi^\ca_g\pm\widehat \psi^\ca_{\bar
d}\rg\cdot \eeq
We remark that $W_{\rm HPS}$ in \Eref{Whi} does not give rise to
mass terms for fermions in the sectors $\uH-\uHb$ and
$\eH-\eHb$. However fermion masses also arise from the
lagrangian terms
\bea \nonumber &-i \sqrt{2} g K_{\al\bbet}\Bigg(
\sum_{a=1}^{15}\ovl \ldu_{\rm C}^a \lf\ovl\psi^\al_{\bar{H}^c} \lf
T^a\bar{H}^{c\trc}\rg^\bbet-\ovl\psi^{\al\trc}_{H^c} \lf T^{a*} H^{c}\rg^\bbet \rg+\\
&\sum_{m=1}^{3} \ovl\ldu_{\rm R}^m \lf\ovl\psi^\al_{\bar{H}^c} \lf
T^m \bar{H}^{c\tr}\rg^\bbet- \ovl\psi^{\al\tr}_{H^c} \lf T^{m*}
H^{c}\rg^\bbet \rg+{\rm h.c.}\Bigg), \label{a4}\eea
where $\ldu_{\rm C}^a$ [$\ldu_{\rm R}^m$], is the gaugino
corresponding to the generator $T_{\rm C}^a$ [$T_{\rm R}^m$] and
$\psi_{\bar{H}^c}$, $\psi_{H^c}$ represent the chiral fermions
belonging to the superfields $\bar{H}^c$, $H^c$ respectively.
Concentrating on $T_{\rm C}^{15}$, $T_{\rm R}^3$, we obtain
\bea \nonumber&& -i \sqrt{2} g \lf \ovl \ldu_{\rm
C}^{15}\lin{\ovl\psi_\nu}{\ovl\psi_{\bar\nu}}{M_K\over
f^2}\stl{-T^{15*}_{\rm C}\Hcc}{T^{15}_{\rm C}{H^{c\trc}}}+\ovl
\ldu_{\rm C}^{15}\lin{\ovl\psi_\nu}{\ovl\psi_{\bar\nu}}{M_K\over
f^2}\stl{-T^{3*}_{\rm R}\Hcc}{T^{3}_{\rm R}{H^{c\tr}}}\rg+{\rm
h.c.}= \\ &&{igh\over2f}
{{\psi}_{\bar\nu}-{\psi}_{\nu}\over\sqrt{2}}\lf-\sqrt{3\over2}\ldu^{15}_{\rm
C}+\ldu^3_{\rm R}\rg+{\rm h.c.}=-im_{\perp}\what
\psi_{\nu-}\ldu^\perp+{\rm
h.c.}\>\>\>\mbox{with}\>\>\>\stl{\ldu^{\perp}}{\ldu^{||}}=U_{\rm
CR}\stl{\ldu_{\rm C}^{15}}{\ldu_{\rm R}^{3} }.\nonumber \eea
Therefore, we obtain a Dirac mass term between the chiral fermion
$\what \psi_{\nu-}$ and the gaugino $-i \ldu^{\perp}$, whereas a
Dirac spinor composed by the combination of $\what \psi_{\nu+}$
and $-i \ldu^{||}$ remains massless and can be interpreted as the
Goldstino which signals the (spontaneous) SUSY breaking along the direction of
\Eref{inftr}.

Similarly, focusing on the directions $T_{\rm C}^{8+2\ca}$ with
$\ca=1,2,3$ and $T_{\rm R}^{1}$ and $T_{\rm R}^{2}$, we obtain the
mass terms
\beq i{gh\over2\sqrt{f}}\lf\sum_{\ca=1}^3\lf
\what\psi^\ca_{u}\ldu^{\ca+}_{\rm C}-\what\psi^\ca_{\bar
u}\ldu^{\ca-}_{\rm C}+{\rm h.c.}\rg+\lf\what\psi_{e}\ldu^-_{\rm
R}-\what\psi_{\bar e}\ldu^+_{\rm R}+{\rm
h.c.}\rg\rg,\label{mpm}\eeq
where we have defined the following combinations of gauginos
\beq \label{lpm} \ldu_{\rm C}^{\ca\pm} ={1\over\sqrt{2}}
\lf\ldu_{\rm C}^{7+2\ca} \pm i \ldu_{\rm
C}^{8+2\ca}\rg\>\>\>\mbox{for}\>\>\>\ca=1,2,3\>\>\>\mbox{and}\>\>\>\ldu_{\rm
R}^\pm = {1\over\sqrt{2}}\lf\ldu_{\rm R}^1 \pm i \ldu_{\rm
R}^2\rg\eeq
in agreement with the definition of the corresponding gauge bosons
in \Eref{Apm}. Therefore, the chiral fermions $\what\psi^\ca_{u}$
and $\what\psi^\ca_{\bar u}$ [$\what\psi_{e}$ and
$\what\psi_{\bar{e}}$] combine with $\ldu_{\rm C}^{\ca\pm}$
[$\ldu_{\rm R}^\pm $] to form two Dirac (or four Weyl) fermions
with mass $m_\pm$ as one deduces from \Eref{mpm}. This
completes the derivation of the spectrum of the model along the
inflationary trajectory of \Eref{inftr}.

\setcounter{equation}{0} \setcounter{subsubsection}{0}
\renewcommand{\theequation}{B.\arabic{equation}}
\renewcommand{\thesubsubsection}{B.\arabic{subsubsection}}
\section*{Appendix B: Inflaton Oscillations After non-MHI}\label{pre}

In this Appendix we discuss various (p)reheating mechanisms
\cite{preheating} which could become competitive with the
perturbative decay of the inflaton to lighter degrees of freedom,
as analyzed in \Sref{pfhi}. Indeed, in certain regions of the
parameter space, the process of reheating in this theory can be
quite complex. After the end of non-MHI, the inflaton develops a
tachyonic mass, crosses an inflection point and enters into a
phase of damped oscillations. As pointed out in \cref{garcia},
where a similar potential is investigated, the particle production
due to tachyonic preheating is not significant because the passage
of the inflaton through this region is very short. During the
subsequent oscillations, perturbative production of superheavy
bosons -- i.e., bosons with masses at the SUSY vacuum proportional
to $\Mpq$ -- is not possible, since these particles are heavier
than the inflaton at the global minima of its potential as shown
in \Tref{tab3}. Therefore effects of narrow parametric resonance
\cite{preheating} are also absent. However, if the initial
amplitude of the inflaton oscillations is large enough, it may
pass through zero, $h=0$, where these bosons are effectively light
and can be produced through (non-perturbative) instant preheating
\cite{instant}, as we discuss in \Sref{pre2} below. We first study
the dynamics of the inflaton' s oscillations in \Sref{pre1}.

\subsubsection{Dynamics of the Inflaton Oscillations  \label{pre1}}

The cosmological evolution of $\what h$ ($h$) in the EF is
governed by the equation of motion:
\beq\ddot{\what{h}}+3\widehat{H}\dot {\what{h}}+\widehat V_{{\rm
HI0},\what{h}}=0\>\>\> \label{eqf}\eeq
where $\He$ is the Hubble parameter in the EF and $\Vhio$ is given
in \Eref{Vhi} -- recall that the dot denotes derivation w.r.t. the
cosmic time $t$ and $h={\sf Re}(\snH+\snHb)/2$ along the direction
of \Eref{inftr}. In the LHS of \Eref{eqf}, we neglect the damping
term $\Gamma_{\rm I}\dot{\what h}$ which is important only at the
stage of rapid oscillations of $\what h$ near one of the minima of
$\Vhio$ \cite{preheating}. Note that contrary to the case of the
potential analyzed in \cref{garcia}, the minima of $\Vhio$ lie at
$|h|=2\Mpq\gg0$ and $\Vhio$ has a maximum at $h=0$ with
\beq\Vhio(h=0)=\what V_0=\ld^2\Mpq^4\label{V0}\eeq
which can not be ignored. Due to these features, the quadratic
approximation to $\Vhio$ is not accurate enough for the
description of the $h$ post-inflationary evolution.

The solution of Eq.~(\ref{eqf}) can be facilitated if we use as
independent variable the number of e-foldings $\Ne$ defined by
\beq \Ne=\ln \left(\widehat R/\widehat R_{\rm
i}\right)~\Rightarrow~\dot \Ne =\He\>\>\mbox{and}\>\>\dot
\He=\He'\He\,. \label{Ndfn} \eeq
Here the prime denotes derivation w.r.t. $\Ne, \widehat R(t)$ is
the EF scale factor and $\what R_{\rm i}$ is its value at the
commencement of non-MHI, which turns out to be numerically
irrelevant. Converting the time derivatives to derivatives w.r.t.
$\Ne$, Eq.~(\ref{eqf}) is equivalent to the following system of
two first order equations
\beq F_h =J \He \what R^3 h'\>\>\mbox{and}\>\> J\He F_h'=
-\Ve_{{\rm HI0},h}\what R^3\>\>\>\mbox{with}\>\>\>F_h =\dot{\what
h}\what R^3. \label{eqfN}\eeq
This system can be solved numerically by taking
\beq \label{Hosc} \He={1\over\sqrt{3}\mP}\lf F_h^2/2R^6+\what
V_{\rm HI0}\rg^{1/2}\eeq
and imposing the initial conditions (at $\Ne=0$)
$h(0)=(0.5-2.5)\mP$ and $h'(0)=0$. We checked that our results are
pretty stable against variation of $h(0)$.

During non-MHI  we have $\He\simeq \He_{\rm HI0}$ and the results
of \eqs{s*}{lan} are well verified. Soon after the end of non-MHI,
we obtain $\He\simeq \He_{\rm HI0}e^{-3\lf\Ne-\Ne_{\rm f}\rg/2}$
-- with $\Ne_{\rm f}$ being the value of $\what N$ at the end of
non-MHI -- and $h$ enters into an oscillatory phase with initial
amplitude equal to $h_{\rm f}$ given by \Eref{sgap}. Since the
value of $\Vhio$ at the end of inflation, $V_{\rm HI0}$ is larger
than the value of $\Vhio$ at its local maximum $h=0$, $\what V_0$,
we expect that $h$ crosses zero at least once during its
evolution. However, as can be deduced from \eqs{sgap}{lan},
lowering $\ld$ increases $h_{\rm f}$ but decreases $\what V_0$.
Therefore the passage of $h$ through zero is facilitated.

\begin{figure}[!t]\vspace*{-.15in}
\hspace*{-.08in}
\begin{minipage}{8in}
\epsfig{file=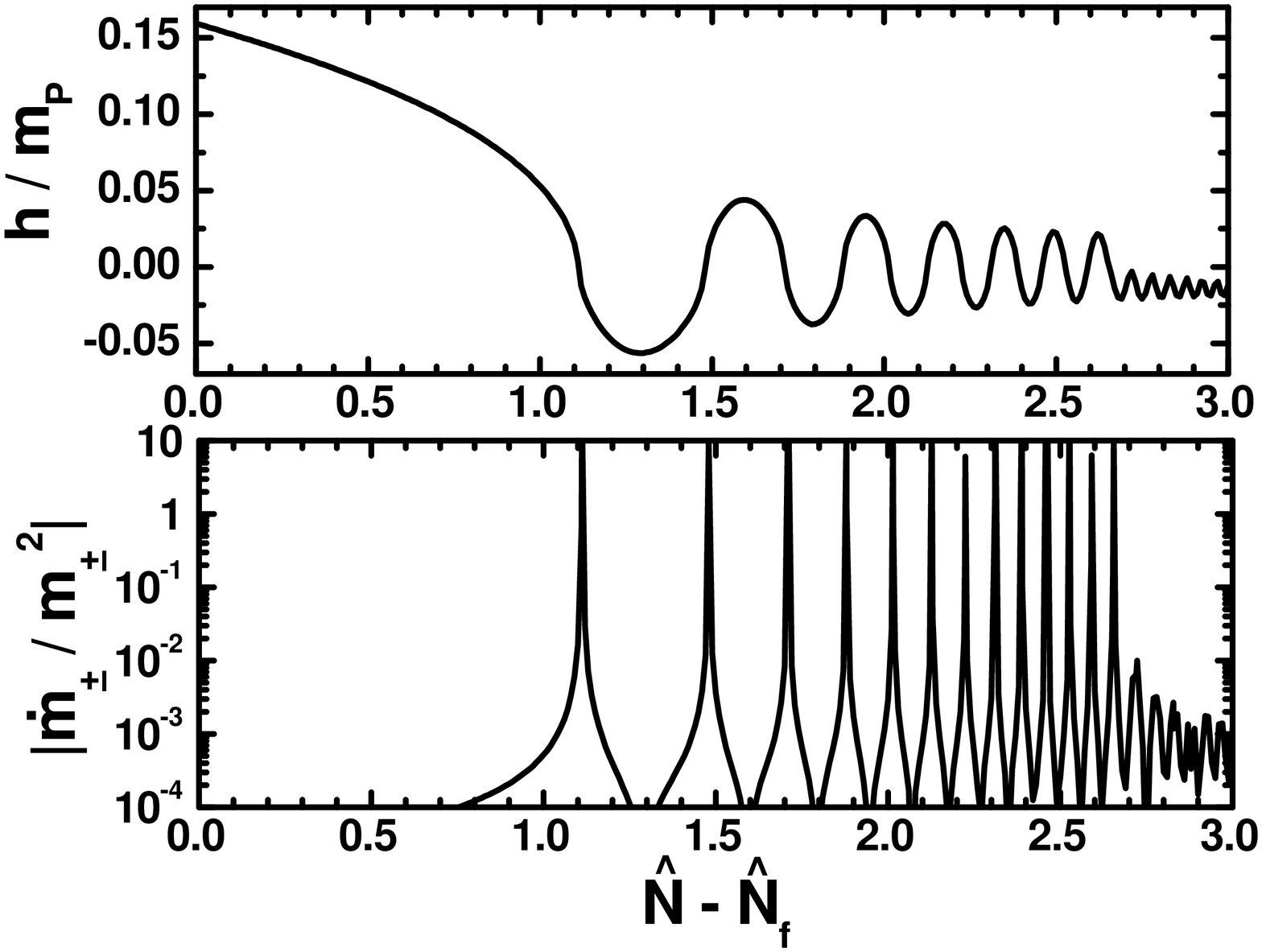,height=3.6in,angle=-90}
\hspace*{-1.25cm}
\epsfig{file=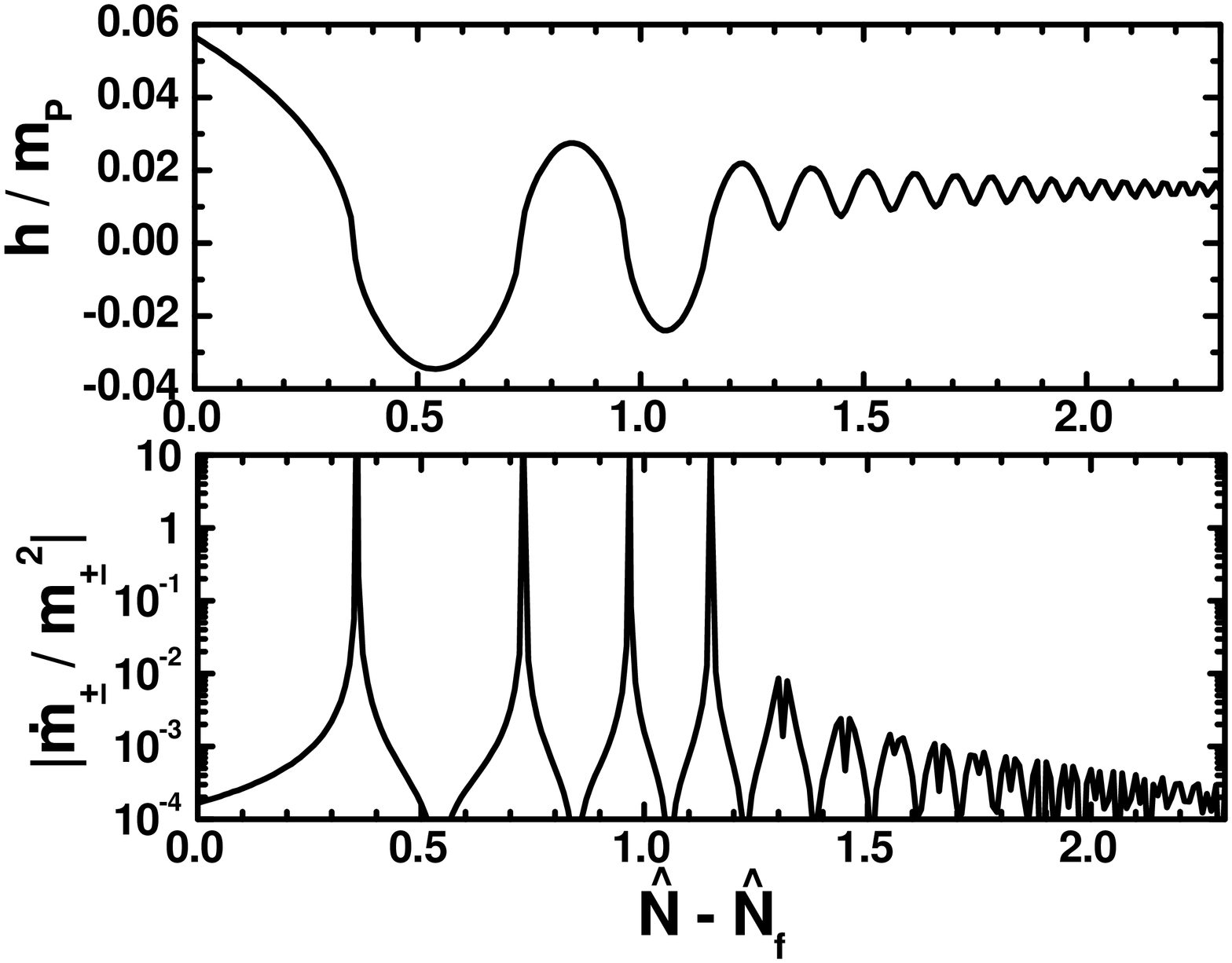,height=3.6in,angle=-90} \hfill
\end{minipage}
\hfill \vchcaption[]{\sl\small  The evolution of the quantities
$h/\mP$ (upper plots) and $|\dot m_\pm/m^2_\pm|$ (lower plots) as
functions of $\what N-\what N_{\rm f}$ for $\ld=0.0037$ and
$\ck=81$ (left panel) or $\ld=0.01$ and $\ck=235$ (right
panel).}\label{fig4}
\end{figure}


The intuitive results above can be established and refined through
the numerical solution of \Eref{eqfN}, during the $h$
oscillations, depicted in the upper plots of \Fref{fig4}. Namely
in left [right] plot we present the evolution of $h$ as a function
of $\what N-\what N_{\rm f}$ for $\ld=0.0037$ and $\ck=81$
[$\ld=0.01$ and $\ck=235$]. In both cases, we see that $h$,
decreasing slowly from its value $h_{\rm f}=0.152$ [$h_{\rm
f}=0.056$] for $\ld=0.0037$ [$\ld=0.01$], passes from the minimum
of $\Vhio$ at $h=2\Mpq$ and then climbs up the hill of $\Vhio$ at
$h=0$, falls towards the other minimum of $\Vhio$ at $h=-2\Mpq$
until it reaches a maximal value and oscillates backwards. This
path is followed some times until $h$ falls finally into one of
the minima of $\Vhio$ at $h=-2\Mpq$ [$h=2\Mpq$] for $\ld=0.0037$
[$\ld=0.01$] -- performing damped oscillations about it. In other
words, $h$ oscillates initially around the local maximum of
$\Vhio$ and then about one of the two SUSY vacua. The number of
passages though zero increases as $\ld$ decreases -- it is equal
to 4 [12] for $\ld=0.01$ [$\ld=0.0037$]. Solving repetitively
\Eref{eqfN} we notice that $h$ ceases to cross $h=0$ for
$\ld>0.088$.

\subsubsection{Instant Preheating \label{pre2}}

Whenever $h$ crosses zero particle production may occur via
instant preheating \cite{instant}. This mechanism is activated
when the $h$-dependent effective masses, $\mef$, of the produced
particles violate the adiabaticity criterion, according to which
\beq \label{adb} \left|\dot{m}_{\rm eff}/\mef^2\right|=\left|\He
\mef'/\mef^2\right|\ll1.\eeq
Here, $\mef$ represents collectively the masses of superheavy
bosons with masses proportional to $g\Mpq$, $\lH \Mpq$ or $\lHb
\Mpq$ -- see \Tref{tab2}. We focus on the production of these
bosons since these can subsequently decay efficiently to the light
SM particles altering drastically the picture of the usual
perturbative reheating. On the contrary, the scalars of the
$S-\snH-\snHb$ sector with masses proportional to $\ld\Mpq$ have
suppressed decay modes to the RH neutrinos only. Taking as an
example $\mef=m_{\pm}$ we plot in the lower plots of \Fref{fig4}
the evolution of $|\dot m_\pm/m^2_\pm|$ as a function of $\what
N-\what N_{\rm f}$ for $\ld=0.0037$ and $\ck=81$ [$\ld=0.01$ and
$\ck=235$] -- see left [right] plot. We observe that \Eref{adb} is
violated more frequently as $\ld$ drops since the passages of $h$
through zero become also more frequent. From the results of our
numerical treatment we find that \Eref{adb} holds during the whole
post-inflationary evolution of $h$ for $\ld>0.045$ whereas it
fails more than 40 times for $\ld<0.001$.

The produced this way superheavy bosons acquire a large mass while
the inflaton increases towards its maximum amplitude and start to
decay into all lighter particles within almost half oscillation of
the inflaton, rapidly depleting their occupation numbers. As shown
in \cref{garcia} an efficient transfer of energy from $h$ to the
superheavy bosons requires a rather large (let say $50-70$) number
of passages of $h$ through zero. Meanwhile effect of backreaction
of the produced particles on the $h$ condensate may become
significant and a more involved numerical study of the process is
imperative. Trying to deliberate our leptogenesis scenario from
such a complicate situation, we impose the indicative lower bound
$\ld\geq0.001$ -- which can be translated as a bound $\ck\geq21$
via \Eref{lan} -- above which our estimations in \Sref{pfhi} are
more or less independent of the preheating effects. We hope to
return to the analysis of $\ld<0.001$ region in the future.


\rhead[\fancyplain{}{ \bf \thepage}]{\fancyplain{}{\sl Non-MHI \&
non-Thermal Leptogenesis in a SUSY PS Model}}
\lhead[\fancyplain{}{\sl \leftmark}]{\fancyplain{}{\bf \thepage}}
\cfoot{}

\end{document}